\documentclass[aps,a4paper,floatfix,nofootinbib]{revtex4}

\usepackage{graphicx,color}
\usepackage{amsmath,amssymb,amsfonts,amsthm,amscd,bm}
\usepackage{booktabs}
\usepackage{cancel}

\def\tilde{\widetilde}
\def\bar{\overline}

\def\*{\star}
\def\[{\left[}
\def\]{\right]}
\def\({\left(}      

\def\){\right)}

\def\zbar{{\bar{z} }}
\def\frac#1#2{\dfrac{#1}{#2}}
\def\inv#1{\dfrac{1}{#1}}
\def\half{\tfrac{1}{2}}
\def\d{\partial}

\def\ket#1{ | #1 \rangle}

\def\2pi{\hbox{$2\pi i$}}

\def\dsl{\raise.15ex\hbox{/}\kern-.57em\partial}
\def\Dsl{\,\raise.15ex\hbox{/}\mkern-.13.5mu D}
      \def\CC{{\cal C}}
      
   \def\CH{{\cal H}}   \def\CI{{\cal J}}
   \def\CK{{\cal K}}   
\def\CM{{\cal M}}   \def\CN{{\cal N}}   \def\CO{{\cal O}}
\def\CP{{\cal P}}   \def\CQ{{\cal Q}}   
\def\CS{{\cal S}}   \def\CT{{\cal T}}

\def\2pi{\hbox{$2\pi i$}}

\def\dsl{\raise.15ex\hbox{/}\kern-.57em\partial}
\def\Dsl{\,\raise.15ex\hbox{/}\mkern-.13.5mu D}
\font\numbers=cmss12
\font\upright=cmu10 scaled\magstep1
\def\stroke{\vrule height8pt width0.4pt depth-0.1pt}
\def\topfleck{\vrule height8pt width0.5pt depth-5.9pt}
\def\botfleck{\vrule height2pt width0.5pt depth0.1pt}
\def\Zmath{\vcenter{\hbox{\numbers\rlap{\rlap{Z}\kern
    0.8pt\topfleck}\kern 2.2pt
    \rlap Z\kern 6pt\botfleck\kern 1pt}}}
\def\Qmath{
    \vcenter{\hbox{\upright\rlap{\rlap{Q}\kern3.8pt\stroke}\phantom{Q}}}}
\def\Nmath{\vcenter{\hbox{\upright\rlap{I}\kern 1.7pt N}}}
\def\Cmath{\vcenter{\hbox{\upright\rlap{\rlap{C}\kern
                   3.8pt\stroke}\phantom{C}}}}
\def\Rmath{\vcenter{\hbox{\upright\rlap{I}\kern 1.7pt R}}}

\def\Xmath{\vcenter{\hbox{\numbers\rlap{\rlap{X}\kern
    0.8pt\topfleck}\kern 2.2pt
    \rlap X\kern 6pt\botfleck\kern 1pt}}}

\def\Z{\ifmmode\Zmath\else$\Zmath$\fi}
\def\Q{\ifmmode\Qmath\else$\Qmath$\fi}
\def\N{\ifmmode\Nmath\else$\Nmath$\fi}
\def\C{\ifmmode\Cmath\else$\Cmath$\fi}
\def\R{\ifmmode\Rmath\else$\Rmath$\fi}
\def\barray{\begin{eqnarray}}
\def\earray{\end{eqnarray}}
\def\beq{\begin{equation}}
\def\eeq{\end{equation}}

\def\kvec{{\bf{k}}}
\def\xvec{{\bf{x}}}

\def\AA{\leavevmode\setbox0=\hbox{h}
\dimen0=\ht0 \advance\dimen0 by-1ex\rlap{\raise.67\dimen0\hbox{\char'27}}A}

\makeatletter
\def\iddots{\mathinner{\mkern1mu\raise\p@
\vbox{\kern7\p@\hbox{.}}\mkern2mu
\raise4\p@\hbox{.}\mkern2mu\raise7\p@\hbox{.}\mkern1mu}}
\makeatother

\theoremstyle{plain}

\theoremstyle{remark}

\def\SU{{\rm SU} (N+1)}

\def\Z{\mathbb{Z}}

\def\K{\CK}

\def\daggerk{{\dagger_{\scriptscriptstyle \kappa}}}
\def\daggerc{\daggerk}

\def\Phitilde{\tilde{\Phi}}

\def\Jtilde{\tilde{J}}

\def\K{\CK}

\def\paulix{\begin{pmatrix} 0 & 1 \\ 1 & 0 \end{pmatrix} }
\def\pauliy{\begin{pmatrix} 0 & -i  \\ i & 0 \end{pmatrix} }
\def\pauliz{\begin{pmatrix} 1 & 0 \\ 0 & -1 \end{pmatrix} }

\def\ketket{\rangle\!\rangle}
\def\brabra{\langle\!\langle}

\def\Phidagger{\Phi^\dagger}

\def\qop{\CQ}

\def\X{{\bf{X}}}

\def\SU2{{\rm SU(2)}}

\def\UV{{\scriptscriptstyle {\rm UV}}}
\def\IR{{\scriptscriptstyle {\rm IR}}}

\def\Ddim{\mathfrak{D}}

\def\smallpsi{{\scriptscriptstyle \psi}}

\def\UV{{\scriptscriptstyle {\rm UV}}}
\def\IR{{\scriptscriptstyle {\rm IR}}}

\def\Ddim{\mathfrak{D}}

\def\SL2Z{{\rm SL}(2,\Z)}

\def\psiplus{\psi_{\plusdot}}
\def\psiminus{\psi_{\minusdot}}
\def\psipm{\psi_{\pmdot}}

\def\Pplus{P_{\plusdot}}
\def\Pminus{P_{\minusdot}}
\def\Ppm{P_{\pmdot}}

\def\Hplus{\CH_{\plusdot}}
\def\Hminus{\CH_{\minusdot}}
\def\Hpm{\CH_{\pmdot}}

\newcommand{\plusdot}{\mathbin{+\mkern-5mu\raise0.42ex\hbox{$\cdot$}}}
\newcommand{\minusdot}{\mathbin{-\mkern-5mu\raise0.42ex\hbox{$\cdot$}}}

\newcommand{\pmdot}{\mathbin{\pm\mkern-5mu\raise0.42ex\hbox{$\cdot$}}}
\newcommand{\mpdot}{\mathbin{\mp\mkern-5mu\raise0.42ex\hbox{$\cdot$}}}

\def\smallpsi{{\scriptscriptstyle \psi}}

\makeatletter
\newcommand{\dotminus}{\mathbin{\mathpalette\dotminus@aux{}}}
\newcommand{\dotminus@aux}[2]{%
  \ooalign{%
    $\m@th#1-$\cr
    \hidewidth$\m@th#1\cdot$\hidewidth
  }%
}
\makeatother

\def\qop{\CQ}

\def\Nanti{\bar{N}}

\def\Nbar{\Nanti}

\begin{document}

\title{Non-perturbative renormalization group for pseudo-hermitian scalar fields  in 4D}

\author{
 Andr\'e  LeClair\footnote{andre.leclair@cornell.edu} 
}
\affiliation{Cornell University, Physics Department, Ithaca, NY 14850,  USA} 

\begin{abstract}

We define  a model of 2 coupled SU(2)  doublets of scalar fields in $4$ spacetime dimensions  which have  a rich  structure of 
renormalization group  (RG) flows  to 1-loop when the SU(2) is broken to U(1).   The model is pseudo-hermitian,   
$H^\dagger = \K H \K^\dagger$ with $\K^\dagger \K = \K^2 =1$,  which makes it  non-unitary,  however in a very specific manner with some desirable properties.        
 We compute the beta functions to 3 loops  from the operator product expansion and show that 
the 1-loop structure of flows persists to higher orders.   For SU(2) broken to U(1),    we conjecture a beta function to all orders. 
The flows can be extended to large coupling using a strong-weak coupling symmetry $g \to 1/g$ of the beta functions.        
One finds a line of fixed points which are  non-unitary conformal field theories in 4 spacetime dimensions that were previously unknown.          We also find massless flows between 2 non-trivial fixed points,  and a regime with a cyclic RG flow,  which is allowed since the model is non-unitary.      For the flows between fixed points on the critical line,   we compute the anomalous dimensions of the perturbations in the UV and IR,   and identify some special points where anomalous dimensions  are rational numbers.

\end{abstract}

\maketitle
\tableofcontents

\section{Introduction}

There is a conspicuous  shortage of well-understood conformal field theories (CFT's) in 4 spacetime dimensions that aren't simply free fields,  as for asymptotically free 
non-abelian gauge theory such as QCD \cite{Wilczek,Gross}.   A well known exception is $\CN = 4$ SUSY Yang-Mills theory.
    For marginal perturbations,  such as many ``standard"  generalizations of $\phi^4$ for the Higgs potential,    the renormalization group  (RG) flows are rather limited in scope in exactly 4 spacetime dimensions.     In fact,  this is at the heart  of the so-called hierarchy problem,  namely that straightforward RG arguments would indicate a much higher mass for the Higgs boson since there is no known ultra-violet  (UV) fixed point. Without a known UV completion of the Higgs sector,   there may even be higher dimension operators that make the theory non-unitary.  
  For scalar fields,  with $\phi^4$ interactions,  using  the epsilon expansion in $D=4- \epsilon$ dimensions,  one can find CFT's in 3 spacetime dimensions such as the Wilson-Fisher fixed points,  which have irrational anomalous dimensions \cite{WilsonFisher}.   However in 4D  there are no such fixed points for conventional quartic scalar field interactions,
   which  motivates searching for generalizations,  and this is the main subject of this article.        
This study will lead lead to the definition of new models of quantum field theory (QFT)  that  scrutinize  the idea that all quantum field theories (QFT's) begin and end at RG fixed point CFT's,  since,   as we will see,  our model also possesses cyclic RG flows in addition to typical flows between fixed points,   which  are allowed since our model is non-unitary.

 By contrast,   in two spacetime dimensions (2D)  one has a large toolkit of {\it algebraic} structures to study CFT's,   in particular the Virasoro algebra and affine Lie algebra symmetry of WZW models 
\cite{BPZ,WittenWZW,KZ}.   For the purposes of this article,   we should  point out  that in the classification of minimal models of CFT based on  only the Virasoro algebra,  one finds many more non-unitary theories than unitary,   all with real eigenvalues of the hamiltonian,  and many of them have important physical applications,  such as the Lee-Yang edge singularity.   
Let us also  mention that there  has recently been renewed and growing interest in non-unitary minimal models  of CFT in 2D and 
RG flows between them \cite{MussardoLG,Tanaka,KlebanovMM,Negro,Ravanini}. 
Here the non-unitarity is manifested in negative norm states,  rather than non-hermiticity of the hamiltonian.         
We would also like to stress that constructions  of the unitary  CFT's often  requires projecting onto  unitary sub-Hilbert spaces of non-unitary theories \cite{FQS}.      For instance,   in the Coulomb gas method one must project the Hilbert space of a free boson to the proper space of the minimal model CFT \cite{DotsenkoFateev},   and this can be described as a BRST procedure \cite{BernardFelder},  or a reduction based on the SU(2) quantum group.      In this way one can describe both the unitary and non-unitary minimal models of CFT in 2D in a unified manner.
      This procedure is similar to the BRST treatment of gauge theories which consistently projects out the  ghosts,  i.e. the negative norm states.    
We should also mention that the unitary CFT's can be understood as coset reductions of current-algebras \cite{GKO},  so that current algebras play a fundamental role in the classification of CFT's.      
It would certainly be nice to have such an algebraic formulation of CFT's in 4D,   and  we hope this article  opens such a door, however narrow it may turn out to be. 
Let us state from the outset that we do not formulate such fully consistent projections onto unitary QFT's  in this article,  however we will show how the non-unitarity is manifested in the scattering theory for our model,  and that 2 to 2 particle scattering is actually unitary,   such that the theory is unitary at low energies below the pair-production threshold.

Based on the above remarks,   the primary physical motivation for this article is to find generalizations of $\phi^4$ interactions in 4 spacetime dimensions that have a much richer pattern of RG flows than the standard ones and others that have been considered in the literature.     This led us to the construction of novel quartic interactions of bosonic fields that have a non-trivial algebraic structure which is embodied in the Operator Product Expansion (OPE),  equation \eqref{JOPE} below,    which we repeat 
here:
\beq
\label{JOPE0}
\lim_{x \to y} \, J^a (x) J^b (y) = - \frac{2 \kappa \delta^{ab} }{16 \pi^4 |x-y|^4 }  -  \frac{i f^{abc}}{4 \pi^2 |x-y|^2 } \, J^c (y) + \ldots  ,
\eeq
where the operators $J^a$ are bilinear in the scalar fields and transform in the adjoint representation of SU(2),  and $f^{abc}$ are the usual structure constants of the 
SU(2) Lie algebra.    
    The algebraic structure of this OPE essentially {\it defines} the model.     It is meaningful for any Lie algebra,   however below we will focus only on SU(2).   The proper mathematical context is the subject of Operator Algebras,  and   our model provides a {\it representation} of the above operator algebra with $\kappa = 1$,     however as we will explain  this representation is non-unitary.      The above OPE  closely parallels current algebra OPE's in 2D conformal field theory,   however to our knowledge it has not been considered before.  The reasons for this will be explained below,   in particular that  the operators $J^a (x)$ are actually non-local.     This Operator Algebra does not arise in  standard generalizations of $\phi^4$ models in the literature, 
       thus our models have no overlap with the latter and this is what makes them  completely novel,   since they cannot be recovered under any known limits of standard $\phi^4$ 
    generalizations that are unitary.

As mentioned above,   the model we define below  that leads to the above OPE is in fact non-unitary,   which clearly indicates its distinction with standard $\phi^4$ generalizations.  
This model was  first defined in  \cite{ALDolls} where RG flows were studied to one loop.\footnote{In that  article \cite{ALDolls} we speculated on applications 
to the Higgs sector of the Standard Model of particle physics,  and even went so far as to suggest that the existence of the 3 families of leptons and quarks may originate from a cyclic RG flow.    However in the present article we leave these speculative issues aside and focus on the non-perturbative RG for the models we define.}     In the present article we mainly  consider the model in 
4 dimensional Euclidean space since we are more interested in the RG flows and  the resulting non-unitary CFT's which are interesting in their own right,  
  and this  is sufficient for potential applications to statistical mechanics in 4 spatial dimensions.     
However for potential  applications to quantum mechanics,  specifically quantum field theory in $3+1$ dimensional Minkowski space,    one must address the non-unitarity of the model since there are 
additional positivity constraints in such a context.            
Our  hamiltonian below  is pseudo-hermitian
\beq
\label{Hdagger}
H^\dagger = \K \, H \, \K^\dagger,    ~~~ \K^\dagger \K =1, ~~~~~ \K^\dagger = \K, ~~~~ \Longrightarrow~~~ \K^2 =1 ,
\eeq
which is to say that $H^\dagger$ is unitarily equivalent to $H$ itself.    
This doesn't necessarily entail an unruly   hornets'  nest  if the hamiltonian has the above  additional  algebraic structure beyond simply the  statement 
that $H^\dagger \neq H$.  
 In  \cite{BL38} hamiltonians satisfying \eqref{Hdagger},   were classified according to $\CC,\CP,\CT$  discrete symmetries,  and 38 universality classes of random  hamiltonians were obtained,   extending the 3-fold classifications of Wigner and Dyson based on time-reversal symmetry,  and  the 10-fold classifications of Altland-Zirnbauer which included particle-hole symmetry \cite{Dyson,AltlandZirnbauer}.\footnote{It is an interesting exercise to
specialize the 38-fold classification in \cite{BL38} to relativistic QFT and we hope to publish such results in the future.}    
  These classes have seen many applications to open quantum systems and their Lindbladians (see for instance  the review \cite{OkumaSato} and references therein).         
 In fact pseudo-hermitian hamiltonians were proposed as consistent extensions of quantum mechanics  long ago by  Pauli \cite{Pauli}.     More recently pseudo-hermitian quantum mechanics has been developed in detail by  Mostafazadeh and others \cite{Mosta1,Mosta2,Mosta3},   and is by now a well-established and highly cited framework.    Pseudo-hermiticity as defined by \eqref{Hdagger} is not the same as $\CP \CT$ symmetric quantum mechanics,   which also has real eigenvalues if the $\CP \CT$ symmetry is unbroken \cite{BenderReview}.      Pseudo-hermitian hamiltonians have several desirable features that parallel the usual quantum mechanics,  such as having real energy eigenvalues,  etc.        These general properties of pseudo-hermitian hamiltonians  we need will be reviewed below,  and then specialized to our model in Section III.

The RG flows for our model were studied to 1-loop  in \cite{ALDolls}.     To this order   
  one finds a  line of ultraviolet (UV) or infrared (IR) fixed points,   flows between such fixed points,     and perhaps more interestingly,   some cyclic RG flows. 
 The main motivation for this article was to study whether this  rich structure of RG flows persists to higher orders in the RG beta functions.       
   Since the cyclic RG flows require flowing to infinite values of the couplings,   it is especially important to study whether this cyclic behavior persists to higher orders in the 
   RG beta functions.            We will show that the 1-loop features do indeed persist  non-perturbatively in a very interesting,  non-trivial  manner.

Of the kinds of RG flows described below,   the cyclic flows are the most exotic and unexpected.     
    Due to various c-theorems and their generalizations to 4D a-theorems,   such flows are sometimes considered as impossible,  
 very rare,  or just mathematical curiosities \cite{ctheorem,CardyCtheorem, Osborn,  JackOsborn,Schwimmer,Komargodski,Polchinski,Klebanov}. 
    Let us thus comment on how our results on cyclic RG  flows  fit into the existing literature,  and are actually consistent with it due to the non-unitarity.                  
In K.  Wilson's pioneering work on the renormalization group (RG),     he attempted to classify  possible RG flow behavior in a model independent way \cite{KWilson}.       Although asymptotic freedom,   which is  essential to understanding QCD \cite{Wilczek,Gross},    was curiously overlooked,   he did point out the possibility of more exotic  limit-cycle behavior,   i.e. a cyclic RG flow.      If the period of the RG flow is $\lambda$,   then the RG flow of the couplings implies 
\beq
\label{glambda} 
g(\ell + \lambda) = g(\ell)
\eeq
where $\ell = \log L $ is the logarithm of the {\it length}  scale $L$,
 such that increasing $\ell$ corresponds to a flow to low energies. 
 This can be interpreted as a discrete rather than continuous conformal symmetry.    
The RG period $\lambda$ is the fundamental  physical parameter of such a flow,  and should be an RG invariant,  as it will be in the models defined in this article.       
In Wilson's last work  with Glazek \cite{GW1,GW2},  motivated by results in nuclear physics 
\cite{nuclear},\footnote{The latter corresponds to a limit cycle in the IR.} 
 they  considered a quantum mechanical model in $D=0+1$ spacetime dimensions where a related signature of a cyclic RG is an infinite sequence of eigenstates 
with scaling behavior 
\beq
\label{ERussianDoll}
E_{n+1}  \approx  e^{\lambda} \, E_n  ~~~~~ n=1,2, 3, \ldots ~~~~ {\rm for ~ n ~ large}.
\eeq
A picturesque description of this behavior are nested  ``Russian Dolls",  i.e. Matryoshka Dolls,     wherein the spectrum  repeats itself  indefinitely in every cycle $\lambda$ as one probes higher energies, again at least up to some cut-off.      In relativistic models,    a Russian Doll RG is such that as one probes ever smaller length scales,   one finds the main structures repeat themselves up to some UV cutoff.   
For a general discussion of signatures of RG limit cycles we refer to the  work \cite{Elliptic}.\footnote{The ``Russian Doll" terminology was coined in \cite{LeClairSierra}.}   
For a review of cyclic RG flows mainly aimed at applications to nuclear physics,   see 
\cite{BraatenHammer}.    
Concurrently with Glazek-Wilson's work,   with D.  Bernard \cite{BLflow} we independently  proposed cyclic RG flows for current-current perturbations in relativistic  $2D$ quantum field theory based on the  higher order beta-functions  proposed in \cite{Gerganov}.      Subsequently,    a many body version of Glazek-Wilson's model was proposed in \cite{LeClairSierraBCS} which is a generalization of BCS  superconductivity with an additional 
coupling which breaks time-reversal symmetry $\CT$,   and also exhibits cyclic RG behavior with the above Russian Doll property,  namely there is an infinite sequence of BCS condensates with the behavior \eqref{ERussianDoll}.    For the current-current perturbations considered in \cite{BLflow},    a relativistic  S-matrix was presented in \cite{LeClairSierra},  referred to as the cyclic sine-Gordon model below,   which had the anticipated infinite sequence of resonances with the Russian Doll behavior \eqref{ERussianDoll}.    Furthermore,  the resulting thermodynamic Bethe ansatz was studied indicating 
oscillations of the thermodynamic c-function in the UV \cite{Roman}.

Let us return to the consistency of our  4D models with the various a-theorems studied in  \cite{ctheorem,CardyCtheorem, Osborn,  JackOsborn,Schwimmer,Komargodski,Polchinski}.       The latter studies often utilize quantum fields coupled to gravity since the focus is on the conformal,  or trace anomaly of the stress-energy tensor.        As we will show below,  if  one is forced to  understand  the flows  for all regimes of the 2 couplings in our model,   then a cyclic regime  is  inescapable. 
One  thus needs to understand how our model 
circumvents certain,  perhaps hidden assumptions behind studies that would seem to rule out cyclic RG flows.  
One obvious answer is that non-unitary theories,  in particular those based on pseudo-hermitian hamiltonians,  were not considered before.     In \cite{Polchinski} for instance,  which relies heavily on the dilaton trick of  Komargodski and Schwimmer \cite{Schwimmer,Komargodski},    the assumption of unitarity is explicitly stated.     
 In addition to  this,  and we believe this is more important,    in order to formulate something like a c-theorem,    one needs   a well-defined  perturbation theory about {\it both} the UV and IR  fixed points  such that it applies to flows between fixed points,  and this was also assumed in \cite{Polchinski}.        For the cyclic regime of our model,   there are no such fixed points about which to perturb.         
One should  also mention that cyclic RG flows were found in a narrow,  non-unitary  region of couplings
for $\phi^6$ theory in $D=3 - \epsilon$ dimensions where $\phi$  is a matrix of scalar
fields \cite{Klebanov}.

Let us summarize the main results and organization of this article.     In the next section we define  the models and discuss their discrete $\CC,\CP,\CT$ symmetries. 
The marginal perturbations of the free CFT are what make the model  non-unitarity,    and we argue that since $\K$ does not commute with charge-conjugation $\CC$,   the model breaks $\CC$,
and also $\CC\CP$.             In Section III  we  address the non-unitarity of the model as a quantum field theory in $3+1$ dimensions.  The non-unitarity is manifested as negative norm states,   and we show that the model is in fact unitary at low energies below the threshold for pair-production. 
 In Section IV  we compute the RG beta functions to 3-loops for the fully anisotropic case based only on the OPE \eqref{JOPE0}.   We   specialize to SU(2) broken to U(1) which involves only 2 couplings.       
 By comparing with the algebraic structures and higher loop integrals for current-current perturbations in 2D,   we present arguments which lead to a proposal for the beta functions to all orders.
       In Section V  we study the RG flows in the resulting 2-coupling parameter space to all orders in the couplings.     This is made possible by a certain RG invariant $Q$ where RG flow trajectories are constant $Q$ contours,    and also a strong-weak coupling duality $g \to 1/g$ of the beta functions.        The resulting picture of RG flows presents itself as a kind of master flowchart  for 2 couplings,   since it  includes new fixed point CFT's,  marginally relevant or irrelevant,   asymptotic freedom,  
massless flows between non-trivial fixed points,   and alas cyclic RG flows. This  flowchart is natural  if one starts with a single coupling,   which has a marginally relevant and irrelevant direction.  Opening up this flow with an additional coupling that preserves this flow along the diagonals  leads to this kind of flowchart.     The latter flowchart arises  rather generally for  Kosterlitz-Thouless flows in 2D.    We refer to the Figures below,
in particular Figure \ref{Flows} for the 1-loop master flowchart,    and its generalization to higher orders in subsequent Figures.        For the cyclic flows,  the RG period $\lambda$ is a simple function of the RG invariant $Q$,   namely
$\lambda = 2 \pi/\sqrt{Q}$.   The existence of a regime of couplings where the RG is cyclic calls into question the commonly accepted paradigm that every QFT begins or ends at a fixed point.    
    By taking one of the couplings to be imaginary,    we  also  find massless flows between different fixed points along the critical line.        The latter are 4D versions of flows between   minimal models in 2D which have received much attention recently \cite{MussardoLG,Tanaka,KlebanovMM,Negro}.    
In the Appendix we review current-current perturbations in 2D  for comparison.

\section{Definition of the models and their discrete $\CC, \CP, \CT$ symmetries} 

\def\htilde{\tilde{h}}

In this Section we define the models as marginal perturbations of  free massless bosonic fields.   The free theory is unitary,   however it is the marginal perturbations that make it non-unitary.        In doing so,   we  will already introduce some aspects of pseudo-hermitian quantum mechanics which will be reviewed in more detail in the next section.     
First we review the standard free theory of the massless free bosons and its discrete symmetries,   which will be important for identifying which symmetries are broken by the marginal perturbations.

\subsection{The free theory and its discrete symmetries}

Our models are defined as certain marginal operator perturbations of a conformal field theory consisting of two scalar field doublets.    
Introduce two  independent $\SU2$ doublets of  complex  bosonic  (non-Grassmannian)  spin-0 fields,  
$\Phi_{i}$ and $\Phitilde_{i}$, ${i =1,2}$
and their usual hermitian conjugates $\Phi^\dagger$, $\Phitilde^\dagger$.      
Under $\SU2$ transformations,   $\Phi (x) \to U \Phi (x)$,   where $U$ is a $2\times 2$  unitary $\SU2$ group matrix acting on the ``i" indices in the above equation and the same for $\Phitilde$.   The symmetry is actually $U(2)$ which allows an additional 
U(1) ``hypercharge".    
The free action is the standard one for free  massless bosonic fields.  
\beq
\label{Sfree}
\CS_{0} = \int d^4 x \( \d_\mu \Phi^\dagger \d^\mu \Phi  +  \d_\mu \Phitilde^\dagger \d^\mu \Phitilde \),
\eeq
where $\Phidagger \Phi = \sum_{i=1,2} \Phidagger_i \Phi_i $ etc.    The free theory has global $\SU2 \otimes \SU2$ symmetry since the fields
$\Phi,\Phitilde$ are independent,   however the interactions introduced below will  break this down to the diagonal  $\SU2$. 
The quantization of this theory  is standard \cite{Peskin}.       Consider the quantization of one component $\Phi = \Phi_i$ for $i=1 ~ {\rm or} ~ 2$,   and  the same applies to 
$\Phitilde$.    
Expand  the field in terms of creation/annihilation operators $a_\pm,  a^\dagger_\pm$: 
\barray
\label{as}
\nonumber 
\Phi (x) &=&  \int \frac{d^3 \kvec}{( 2 \pi)^{3/2}} \inv{\sqrt{2 \omega_\kvec}} \( a^\dagger_- (\kvec ) e^{- i k\cdot x}  + a_+(\kvec) e^{i k\cdot x } \) 
\\ 
\Phi^\dagger  (x) &=&  \int \frac{d^3 \kvec}{( 2 \pi)^{3/2}} \inv{\sqrt{2 \omega_\kvec}} \( a_-(\kvec) e^{i k\cdot x}  + a^\dagger_+ (\kvec) e^{-i k\cdot x } \), 
\earray 
where 
$ k \cdot x = \omega_\kvec t - \kvec \cdot \xvec $ with $\omega_\kvec = |\kvec| $,  and similarly for $\Phitilde$.       
Canonical quantization of the  field  $\Phi$ as a boson yields 
\beq
\label{canonical}
[a_\pm (\kvec), a^\dagger_\pm (\kvec')] = \delta^{(3)}  (\kvec - \kvec' ), ~~~~[a_\pm (\kvec), a^\dagger_\mp (\kvec')]  =0, 
\eeq
and the free  hamiltonian is
\beq
\label{freeH}
H_{0}  = \int  d^3  \kvec  \, \omega_{\kvec} \(   a^\dagger_+ (\kvec) \, a_+(\kvec) + a^\dagger_-(\kvec) \, a_-(\kvec)  \). 
\eeq

The above free theory is necessarily  invariant under $\CC\CP\CT$,   which we now review in order to make certain points below.
Consider first charge conjugation implemented by the operator $\CC$:   
\beq
\label{Cconj2}
\CC \CC^\dagger = 1, ~~~ \CC = \CC^\dagger, ~~~ \CC^2 =1. 
\eeq 
 On the $a, a^\dagger$  operators it acts as 
 \beq
\label{ChargeC}
\CC a_\pm(\kvec) \CC = a_\mp (\kvec), ~~~\CC a^\dagger_\pm (\kvec) \CC = a^\dagger_\mp (\kvec).
\eeq
This implies 
\beq
\CC \Phi (x) \CC = \Phi^\dagger (x)\, ~~~~ \Longrightarrow ~~~~~~\CC H_0 \CC = H_0= H^\dagger_0. 
\eeq
Based on  $\CC$  we  identify  {\it particles}  as those created by $a^\dagger_+$ and {\it anti-particles}  as created by $a^\dagger_-$. 
This operator $\CC$  will play a central role below since it will be broken by interactions.  
 
Let us also review the standard  parity  $\CP$ and time-reversal symmetry $\CT$. 
We take them to also satisfy constraints such as in \eqref{Cconj2}.    
Since parity flips the sign of momentum: 
\beq
\label{Paritydef}
\CP  a_\pm (\kvec) \CP =  a_\pm (-\kvec),  
~~~\CP  a^\dagger_\pm(\kvec) \CP =  a^\dagger_\pm (-\kvec), ~~~~~~\Longrightarrow ~~~~\CP \Phi(t,\xvec) \CP = \Phi(t, - \xvec), ~~~
\CP H_0 \CP = H_0.
\eeq
Time-reversal also flips the sign of $\kvec$ and only differs from $\CP$  in that it is anti-unitary, namely $\CT z \CT = z^*$ for $z$ a complex number.   This leads to 
\beq
\label{TimeRev} 
\CT  a_\pm(\kvec) \CT = a_\pm (-\kvec), ~~~\CT  a^\dagger_\pm  (\kvec)\CT = a^\dagger_\pm (-\kvec)~~~~
\Longrightarrow ~~~ \CT \Phi (t, \xvec) \CT = \Phi (-t, \xvec ), ~~~~~ \CT H_0 \CT = H_0.
\eeq
This implies 
\beq
\label{PT}
\CP\CT  a_\pm(\kvec) \CP \CT = a_\pm (\kvec) ,
\eeq
and
\beq
\label{PT2}
\CP\CT  H_0  \CP \CT = H_0.
\eeq

The full  Hilbert space $\CH$,   which diagonalizes the free hamiltonian $H_0$,  consists of multi-particle  states $|\psi\rangle$ of the form 
\beq
\label{states}
\CH = \{ |\psi \rangle \} , ~~~ | \psi \rangle =  |(\kvec_1, s_1), (\kvec_2, s_2), \ldots  (\kvec_n,  s_n)  \rangle \equiv  a^\dagger_{s_1} (\kvec_1 )  a^\dagger_{s_2} (\kvec_2 ) \cdots a^\dagger_{s_n} (\kvec_n ) |0\rangle  ~~~~~
s_i \in \{\pm \}.
\eeq
In $\CH$ all states have a positive norm such that 
\beq
\label{freenorm}
\langle \psi' |\psi\rangle =  \delta_{\smallpsi'  \smallpsi}  >0,   ~~ \Longrightarrow  ~~~   \sum_\smallpsi   |\psi\rangle \langle \psi | = 1,
\eeq
where $\delta_{\smallpsi' \smallpsi}$  symbolically contains overall energy-momentum conserving  Dirac delta-functions.

\subsection{The marginal perturbations}

In defining the marginal perturbations of interest,   it's necessary to introduce an additional  discrete symmetry operator $\K$,   satisfying 
\beq
\label{Kdef}
\K a_\pm (\kvec)  \K = \pm a_\pm  (\kvec) ,  ~~~ \K a^\dagger_\pm (\kvec) \K  = \pm a^\dagger_\pm (\kvec). 
\eeq
The operator $\K$ is unitary: 
\beq
\label{KKdag}
\K^\dagger \K =1,     ~~~~\K = \K^\dagger, ~~~~ \Longrightarrow ~~~ \K^2 =1,
\eeq
and the free hamiltonian is $\K$ invariant:
\beq
\label{HoInv}
\K H_0 \K = H_0 = H^\dagger_0.
\eeq
Every state $\psi$ in the Hilbert space defined above,  spanned by states of the form \eqref{states},  has  a well defined $\K$ eigenvalue
\beq
\label{Kpsi}
\K |\psi\rangle = \K_\smallpsi  \, |\psi\rangle, ~~~~~  \K_\smallpsi \in \{ \pm 1\}, ~~~~~ \K |0\rangle = |0\rangle.
\eeq

The operator $\K$ does not act locally on the fields $\Phi$,  unlike $\CC, \CP, \CT$.    
It  can be expressed in terms of the $U(1)$  charge $\qop$ operator.  
Let $N_\pm$  denote the number operators for particles created by $a^\dagger_\pm$:
\beq
\label{Nops}
N_\pm = \int d^3 \kvec ~ a^\dagger_\pm (\kvec)  a_\pm (\kvec).
\eeq
Then the U(1) charge is 
\beq
\label{qopeqn}
\qop = N_+ - N_-  
\eeq
and one has 
\beq
\label{charges}
 [\qop,  a_\pm] = \mp a_\pm , ~~~~[\qop,  a^\dagger_\pm] = \pm a^\dagger_\pm. 
 \eeq
One can then identify 
\beq
\label{KU1}
\K = e^{i \vartheta \Nanti} ~~ {\rm with} ~  \vartheta = \pi ~ {\rm and} ~  \Nanti \equiv N_- ~~~~ \Longrightarrow~~ [\qop,  \K] = 0,
\eeq
where $\Nanti$ counts the number of anti-particles.     
Clearly,   $\K^\dagger K = 1$  and  $\K^2 = 1$  since  $\Nanti$ is an integer on the Hilbert space.  
 The $\K$ operator does not commute with charge conjugation $\CC$.   One has
 \beq
 \label{CK}
 \CC \K = \K \CC \, e^{i \pi \qop},
 \eeq
 where we have also used that  $\qop$ is an integer on the Hilbert space,
 and 
 \beq
 \label{CQ}
 e^{i \pi \qop} \CC = \CC e^{i \pi \qop}  .
 \eeq
 On 1-particle states where $\qop = \pm 1$,  eq.  \eqref{CK} implies that $\K$ anti-commutes with $\CC$.

Our models will involve $\Phi^2 \Phitilde^2 $ marginal  interactions involving the $\K$-conjugate fields $\Phi^\daggerk$
and $\Phitilde^\daggerk$:
\beq
\label{phidagc}
\Phi^\daggerc  (x) \equiv  \K \, \Phi^\dagger  \K =  \int \frac{d^3 \kvec}{( 2 \pi)^{3/2}} \inv{\sqrt{2 \omega_\kvec}} \(-  a_-(\kvec) e^{i k\cdot x}  + a^\dagger_+ (\kvec) e^{-i k\cdot x } \).
\eeq
Since the operator $\K$ acts non-locally on the fields $\Phi^\dagger$,    the fields $\Phi^\daggerk (x)$ are {\it non-local}. 
For reasons explained in the next section,   define a modified $\K$-inner product which includes $\K$ insertions: 
\beq
\label{Knorm}
\brabra \psi' | \psi \ketket \equiv \langle \psi' | \CK | \psi \rangle = \K_\smallpsi \, \delta_{\smallpsi'\smallpsi}
\eeq
Using $a_\pm | 0 \rangle =0$, due to the extra minus sign in the above equation  one finds the free field 2-point correlation functions for components $\Phi_i$:
\beq
\label{onepoint}
\brabra  \Phi^\daggerc_i  (x) \Phi_j (y)  \ketket = - \brabra  \Phi_i (x) \Phi_j^\daggerc  (y) \ketket  = \frac{\delta_{ij}}{4 \pi^2 |x-y|^2 },
\eeq
where we have used $\K |0\rangle  = |0\rangle$. 
These correlation functions are consistent with $\K$-conjugation:
\beq
\label{Kconj}
\brabra  \Phi^\daggerc (x) \Phi (y)  \ketket^* = \brabra \(  \Phi^\daggerc (x)  \Phi (y)  \)^\daggerk \ketket =  \brabra  \Phi^\daggerc (y) \Phi (x) \ketket, 
\eeq
which allows the extra minus sign in \eqref{onepoint}.    The two-point correlation functions in \eqref{onepoint} are also consistent with causality.  
The above correlation functions,  or propagators,   are {\it implicitly}  time-ordered.    Thus within a time ordered correlation function one has the following OPE:
\beq
\label{Wick1}
\lim_{x \to y} \, \Phi_i^\daggerk (x)  \Phi_j (y) = - \lim_{x \to y} \, \Phi_i (x)  \Phi_j^\daggerk  (y)  = \frac{\delta_{ij}}{4 \pi^2 |x-y|^2 }.
\eeq
where time ordering is implicit.  (See \cite{Peskin} for detailed explanations.) 
Note that the order of operators matters,   and the above OPE is more typical of a fermion,   even though $\Phi$ was quantized as a boson.
This is a manifestation of the non-locality of the operators $\Phi^\daggerk$.

Let us now turn to adding interactions.     As for $\Phi^\daggerk$ above,   
for any operator $A$ define it's $\K$-conjugate as 
\beq
\label{Adaggerk}
A^\daggerk \equiv  \K \, A^\dagger \K .
\eeq
The operator $ \Phi^\daggerc \Phi$ is not its own hermitian conjugate,   but is invariant under $\daggerk$ conjugation:
\beq
\label{biConj}
( \Phi^\daggerc \Phi)^\daggerk  = \Phi^\daggerc \Phi .
\eeq
Motivated by this,   define the operators
\beq
\label{J}
J^a  =   \Phi^\daggerc  \sigma^a \Phi  \equiv \sum_{i,j}  \Phi^\daggerc_i  \sigma^{a}_{ij} \Phi_j ,
\eeq
where $\sigma^a = (\sigma^a)^\dagger$ are the usual hermitian Pauli matrices:
\beq
\label{tauasigma} 
\sigma^1 = \paulix,  ~~~\sigma^2 =  \pauliy,  ~~~\sigma^3 = \pauliz, ~~~{\rm Tr} ( \sigma^a \sigma^b) = 2 \delta^{ab} .
\eeq
and similarly for $\Jtilde^a$.    These operators are invariant under $\K$-conjugation:
\beq
\label{Jconjugation}
\(J^a\)^\daggerk = J^a, 
\eeq
and the same goes for $\Jtilde^a$.   
 The operators $J^a$,  as defined this way,    have  an interesting operator product expansion.    Using the Wick expansion based on 
 \eqref{onepoint} one finds  
\beq
\label{JOPE}
\lim_{x \to y} \, J^a (x) J^b (y) = - \frac{2 \kappa \delta^{ab} }{16 \pi^4 |x-y|^4 }  -  \frac{i f^{abc}}{4 \pi^2 |x-y|^2 } \, J^c (y) + \ldots  ~~~({\rm with} ~ \kappa =1),
\eeq
where the structure constants $f^{abc}$ are 
\beq
\label{commutator}
[\sigma^a, \sigma^b ] = i f^{abc} \sigma^c,  ~~~~~   f^{abc} =2 \epsilon^{abc},
\eeq
with  $\epsilon^{abc}$  the completely anti-symmetric tensor with $\epsilon^{123} = 1$.    
In the above equation $\kappa=1$,  however we introduced $\kappa \neq 1$  for  book-keeping reasons  that will become clear below,  as the $\kappa$ order counts loops for contributions to the beta function.  
In obtaining the above OPE \eqref{JOPE},    the term where the two outermost fields are Wick contracted gives the term
\beq
\label{JOPE2}
 \lim_{x \to y}  \sum_{i,j,k} \,    \inv{4 \pi^2 |x-y|^2} ~\Phi_j (x) \Phi_k^\daggerk (y)  \sigma^{b}_{ki} \sigma^{a}_{ij}  =     \inv{4 \pi^2 |x-y|^2} ~ \Phi^\daggerk (y) \sigma^b \sigma^a \Phi (y), 
\eeq
where we have used the fact that $\Phi$ is a bosonic (non-Grassmann) field to exchange the order of 
$\Phi_j (x) \Phi_k^\daggerk (y)$ for $j \neq k$.    When $j=k$,  the singularity is already contained the in $\kappa$-term in the OPE.  
This prescription  is necessary for the above OPE to close on the operators $J^a (x)$.    

  The above OPE \eqref{JOPE} is central to the remainder of this article.   
For $\SU2$ there is no sum over ``$c$" in the equation \eqref{JOPE} and this will simplify some calculations below,    however this OPE is valid for other Lie algebras.     
The extra minus sign for $\Phi^\daggerk$ in  \eqref{phidagc} is what leads to the commutator in \eqref{commutator} and the resulting structure constant $f^{abc}$ in the OPE \eqref{JOPE}.   If $J^a$ was instead defined as $\Phi^\dagger \sigma^a \Phi$,   it would not satisfy the OPE in \eqref{JOPE}.    Nor would this OPE be satisfied if
$J^a = \Psi^\dagger \sigma^a \Psi $ where $\Psi$ is a  fermionic, i.e.  Grassmann field.  
As we will see,   the beta functions can be determined from the above OPE \eqref{JOPE} since it is valid inside time-ordered correlation functions.       
Let us emphasize that since we used the Wick expansion in obtaining the OPE in \eqref{JOPE},  it should be understood as an operator equation inside a time-ordered correlation function. 

Since $J^a (x)$ are quantum operators,  they do not necessarily commute.  If one simple exchanges $a,b$ and simulaneously $x,y$ on the LHS of \eqref{JOPE},  then it becomes 
$J^b(y) J^a (x)$  which is not equal to $J^a(x) J^b (y)$ according to the RHS of equation \eqref{JOPE} because of the anti-symmetry of the structure constant $f^{abc}$:
\beq
\label{notcommute}
  \lim_{x \to y} \, J^a (x) J^b (y)  \neq  \lim_{y \to x} \, J^b (y) J^a (x) .
  \eeq
 The above implies that the operators $J^a (x)$ are non-local,   which,  as explained above,    is a consequence of $\Phi^\daggerk$ being non-local.   
 
What is meaningful is that the operator equation \eqref{JOPE} is consistent with $\K$-conjugation,   and from this one can see why the  $i= \sqrt{-1}$ is necessary.    Taking the
$\K$-conjugate of both sides of \eqref{JOPE} one finds
\beq
\label{JOPEconj}
\( J^a (x) J^b (y) \)^\daggerk = J^b(y) J^a (x) = - \frac{2 \kappa \delta^{ab} }{16 \pi^4 |x-y|^4 }  +  \frac{i f^{abc}}{4 \pi^2 |x-y|^2 } \, J^c (y) ,
\eeq
and this is identical to  \eqref{JOPE} since $f^{abc} = - f^{bac}$.   

For euclidean QFT,      the perturbing operators of a CFT are normally  required to be local for physical reasons. 
   Below,   the marginal perturbations will be  
$\CO^a (x) =   J^a (x) \Jtilde^a (x)$  which are indeed local as is evident from the OPE in equation \eqref{OOPE} below.   
To help in  clarifying  this,    it is useful to make an analogy with free massless scalar fields in 2 spacetime dimensions,  especially in light of the comparison 
with 2D current-current perturbations made in the Appendix.     In 2D,  the conserved current operators are in fact local,   however here the operators $J^a$ are not local,  nor are they  conserved currents.   
Consider a free massless scalar field in 2D with the local propagator $\langle \phi (x) \phi (y) \rangle = - \log |x-y|^2$.   In 2D CFT,  the field $\phi (x)$ is 
separated into left and right moving parts:   $\phi(x) = \varphi (z) + \bar{\varphi} (\bar{z})$  where $z = x + i y, ~ \bar{z} = x- iy$.   The two-point function 
is $\langle \varphi (z) \varphi (w) \rangle = - \log (z-w)$ and similarly for $\bar{\varphi}$. 
The field  $\phi$  is local,   but $\varphi$ and $\bar{\varphi}$ are non-local.    Consider a perturbation by  local operators $e^{i \alpha \phi} = e^{i \alpha \varphi} e^{i \alpha 
\bar{\varphi}} $.   Then the OPE of the holomorphic factor is non-local.   Using the Wick expansion,    
\beq
\label{scalarope}
\lim_{z \to w} \,  e^{i \alpha \varphi (z) } e^{i \alpha \varphi (w) }  \sim (z-w)^{\alpha^2}  \, e^{2 i \alpha \varphi (w) } .
\eeq
The above OPE is obviously non-local in that it is not invariant under the exchange of $z$ and $w$,  analogous to the OPE \eqref{JOPE}.  
On the other hand the OPE  of  the local field $e^{i \alpha \phi (x)}$ with  $e^{i \alpha \phi (y)}$  is proportional to $|x-y|^{2 \alpha^2}$ which is local.

These are all the ingredients necessary to define our models.    Define the marginal operators 
\beq
\label{OAdef} 
\CO^a (x) =   J^a (x) \Jtilde^a (x) .
\eeq
The models are then defined by the action 
\beq
\label{perturbations}
\CS = \CS_{0}  + 2 \pi^2  \int d^4 x \,  \sum_{a=1,2,3}  g_a  \, \CO^a (x), 
\eeq
where $g_a$ are 3 independent couplings.   
The factor of $2 \pi^2$ is introduced to simplify the beta-functions of the next section,  such that no 
$\pi$'s appear.      
When $g_1 = g_2 = g_3$ the model has SU(2) symmetry since it is built on the quadratic Casimir.    
The operators $J^a$ are not hermitian but pseudo-hermitian as stated in \eqref{Jconjugation}.  
Thus the  interacting hamiltonian is pseudo-hermitian:
\beq
\label{Hdag2}
H^\dagger = \K H \K  ~~~ \Longrightarrow ~~ H^\daggerc = H .
\eeq

 \subsection{$\CC, \CP$ and $ \CT$ for the interacting theory:   breaking of  $\CC \CP$}

The free theory with hamiltonian $H_0$ is invariant under separate $\CC, \CP, \CT$ symmetries thus respects the $\CC\CP\CT$  theorem since  it does not break the combination 
$\CC\CP\CT$.      Since the proof of the $\CC\CP\CT$  theorem assumes hermiticity and unitarity,     $\CC\CP\CT$  is actually  broken for the interacting theory.      Since 
$\CC$ exchanges 
$a^\dagger_\pm$ according to \eqref{ChargeC},   and $\K$ distinguishes between particles and anti-particles,    then it is clear that $\CC$ is broken.   
This is manifested in the fact that $\CC$ does not commute with $\K$,    as \eqref{CK} shows.   On the other hand,  based on \eqref{PT},   $\K$ does commute with
$\CP\CT$,  so the breaking of $\CC\CP\CT$ is due only to the breaking of $\CC$.      Furthermore,   since $\K$ commutes with both $\CP$ and $\CT$ separately,   then 
$\CP$ is unbroken.     Thus our model breaks $\CC\CP$.\footnote{More generally,  if  $\CC\CP\CT$  is broken,   then this can  allow  for  violations of the spin-statistics theorem since the 
$\CC\CP\CT$ theorem assumes a hermitian hamiltonian and unitarity.         This is not relevant for this article where we quantized the scalar field
$\Phi$ as a boson.       However the  pseudo-hermitian properties of our model have very strong parallels with that
of symplectic fermions,   where a scalar field can be consistently quantized as a fermion \cite{LeClairNeubert, CYLeeSpinStat}.}

\def\htilde{\tilde{h}}

\section{Manifestations of non-unitarity:   negative norm states and a low energy  unitary regime}

As stated in the Introduction,   the  main focus of this article is on the CFT's and RG flows between them for our model,  
and this is  as well-motivated as the extensive literature on flows between unitary and non-unitary models of CFT in 2D,  which we reviewed in the Introduction. 
Our model is strictly speaking non-unitary since $H^\dagger \neq H$,   and this cannot be completely ``fixed"  by the considerations  in this article.    
   However pseudo-hermiticity,  $H^\daggerk = H$,    provides some additional  algebraic structure with some desirable properties which we review in this section.   
  For potential applications to  relativistic particle  physics in $3+1$ spacetime dimensions,     a consistent unitary quantum mechanics would  of course be desirable.    
 The classification of  pseudo-hermitian hamiltonians in \cite{BL38} has led to many applications to {\it open} quantum systems and their Lindbladians. 
  Let us mention that  whether the Universe as a whole is open or closed is not a settled issue.  For instance,   the UV completion of the Standard Model could involve higher dimensional operators that are pseudo-hermitian and thus break unitarity.             
  Let us also mention that it has been proposed that for a potential dS/CFT correspondence,   the CFT is expected to be non-unitary \cite{Strominger},   and 
symplectic fermions have been considered in this context \cite{LeClairNeubert}.       As we will see,   there are very strong parallels between the  pseudo-hermitian  bosonic scalar model we study in this article  and self-interacting symplectic fermions since the non-unitarity is of a very similar nature.
  The non-unitarity of symplectic fermions has been studied in 
several articles \cite{Chengdu,Ryu,CYLee,CYLeeSpinStat} where it has been argued that the interacting model has a consistent generalized unitary time evolution  at least  for 2-body to 2-body scattering \cite{CYLee}.

In this section we study how the non-unitarity of our  model is specifically manifested,   in the S-matrix for instance.         
The subject of Quantum Mechanics for pseudo-hermitian hamiltonians is by now well developed,    with extensive reviews 
\cite{Mosta1,Mosta2,Mosta3}.   These works show that in principle one can still define a unitary theory from a pseudo-hermitian hamiltonian \cite{Mosta1},  however  the procedure for doing so is model dependent and we could not apply it directly to our theory.   
As we will explain,    the manner is which the non-unitarity is manifested in our model  will lead us to conclude that,  although our model is  strictly speaking non-unitary,
 it is in fact  perfectly unitary at low enough energies,   specifically below the energy  threshold for pair-production.    Thus our model provides an example where a non-unitary theory at high energies can still be unitary at low energies.     Similar examples in 2D are discussed below wherein a non-unitary perturbation can still lead to RG flows between unitary CFT's (see Section VB).

To set the stage,  let us first review the  Born rule for probabilities in standard unitary quantum mechanics in the Schr\"odinger picture.    
Assume we are given a Hilbert space $\CH$ with a conventional positive definite inner product.   
Namely for  $|\psi \rangle \in \CH$,     there exists an inner product such that $\langle \psi' | \psi \rangle$ is positive definite.   Indeed,  the  states in \eqref{states} comprise such a Hilbert space.       
At time $t=t_0 =0$ the states need to be properly normalized such that $\langle \psi | \psi \rangle = 1$.      
There exists a basis in $\CH$,     $|\psi_n\rangle \in \CH$ where for simplicity of notation we assume $n$ is discrete,   
$n = 1, 2, \ldots \infty$.     Then 
\beq
\label{psins}
|\psi \rangle = |\psi (t=0) \rangle = \sum_n  c_n \,  |\psi_n \rangle,      ~~~~\langle \psi_m | \psi_n \rangle = \delta_{nm} ,
\eeq
for some complex numbers $c_n$.     
Then 
\beq
\label{norm1} 
\langle \psi | \psi \rangle = 1  ~~~~~~\Longrightarrow ~~~ \sum_n  c_n^* c_n = 1 ,
\eeq
where $c_n^*$ denotes complex conjugation.    According to the Born rule,   $|c_n|^2$  represents the probability that $|\psi \rangle$ is measured to be in the state $|\psi_n\rangle$.     Given a hamiltonian $H$,  or any other operator  $A$ on $\CH$,   then $A^\dagger$ is defined in the standard way with respect to the above inner product.     
Let us now turn to pseudo-hermitian hamiltonians in general.

\subsection{General properties of pseudo-hermitian Quantum Mechanics}
 
Pseudo-hermitian hamiltonians have many desirable properties that parallel those in ordinary hermitian quantum mechanics,  and we here list the most important
for such a theory generally,  i.e. properties that are independent of the model.

\bigskip

\noindent (i) ~~{\it Modified indefinite  metric  on the Hilbert space}
\medskip

Suppose the hamiltonian operator $H$ is pseudo-hermitian with the structure in \eqref{Hdagger},  which we repeat here:
\beq
\label{HdaggerAgain}
H^\dagger = \K \, H \, \K^\dagger,    ~~~ \K^\dagger \K =1, ~~~~~ \K^\dagger = \K, ~~~~ \Longrightarrow~~~ \K^2 =1 .
\eeq
In other words,   $H^\dagger$ is unitarily equivalent to $H$.  
This leads us to define a new inner product which includes an insertion of the operator $\CK$:
\beq
\label{Knorm}
\brabra \psi' | \psi \ketket \equiv \langle \psi' | \CK | \psi \rangle = \K_\smallpsi \, \delta_{\smallpsi'\smallpsi}
\eeq
In this new inner product,    kets have an accompanying $\K$ but bras do not.    
In general,    this new inner product  can have  negative norm states $|\psi\rangle$ if $\K_\smallpsi$ is negative.   
This implies negative probabilities since \eqref{norm1} becomes 
\beq
\label{negprob}
 \sum_n  \K_n \, |c_n|^2 = 1.
 \eeq

\bigskip

\noindent (ii) ~~{\it  Pseudo-hermitian operators correspond to observables with real eigenvalues}
\medskip

For any operator $A$ on $\CH$,     define its pseudo-hermitian  conjugate as follows:
\beq
\label{pseudodag}
A^\daggerk \equiv \CK \, A^\dagger  \, \CK .
\eeq
We define  a pseudo-hermitian operator $A$ as one that satisfies $A^\daggerk = A$.   
For our model,    the  hamiltonian is  such a pseudo-hermitian operator:
\beq
\label{pseudoH}
H^\daggerk = H .  
\eeq
$A^\daggerk$ is the proper conjugation based on the $\CK$-metric in \eqref{Knorm},   namely
\beq
\label{Astar}
\brabra \psi' | A | \psi \ketket^* =  \brabra \psi | A^\daggerk  | \psi' \ketket ,
\eeq
where $*$ denotes ordinary complex conjugation.    
From \eqref{Astar},   one concludes that pseudo-hermitian operators,   in particular the hamiltonian H,  has real eigenvalues. 
In other words,   $H$ is effectively hermitian but on a Hilbert space with indefinite norm.   
Any operator satisfying $A^\daggerk = A$ in principle can correspond to an observable.   
One can easily establish that this pseudo-hermitian adjoint satisfies the usual rules, e.g.\
\barray\label{rules1}
   (AB)^\daggerc &=& B^\daggerc A^\daggerc \,, \nonumber\\
   (a A + b B)^\daggerc &=& a^* A^\daggerc + b^* B^\daggerc \,,
\earray
where $A,B$ are operators and $a,b$ are complex numbers.

\bigskip

\noindent (iii) ~~{\it  Conservation of probability}
\medskip

 The hamiltonian determines the time evolution of  a state $|\psi \rangle =|\psi (t=0)\rangle$,    namely 
 $| \psi (t) \rangle =  \,e^{-i H t} |\psi \rangle$.       By conservation of probability we mean that the norm of states is preserved under time evolution.     More generally:
 \beq
 \label{timeevolution}
 \brabra \psi' (t) | \psi (t) \ketket = \langle \psi'| e^{i H^\dagger t } \, \K \, e^{-i H t} \, | \psi \rangle 
 = 
 \langle \psi'| \K e^{i H t }  \K^2  \, e^{-i H t} \, | \psi \rangle 
 = 
 \brabra \psi' (0) | \psi (0) \ketket 
 \eeq
 In general,   the modified $\K$-metric $\brabra \psi' | \psi \ketket$ introduces negative norm states which imply negative probabilities.  The above statement of unitarity just states that the sum of these probabilities is maintained under time evolution,
 but one must still deal with negative norm states due to the negative probailities.

 \bigskip
 
 \noindent (iv) ~~{\it  Projection operators}
\medskip

The properties of $\K$ naturally lead to projection operators onto positive or negative norm states.      
 Since $\K^2 =1$,   states can be classified according to $\K = \pm 1$.     Namely,
 \beq
 \label{Kvalues}
 \K | \psi_n \rangle = \K_n  \, | \psi_n \rangle,   ~~~~  \K_n \in \{\pm 1 \} .
 \eeq
 Define the operators\footnote{We adopt the notation $\pmdot$ to refer to $\K=\pm$ norm states in order to avoid confusion with $\pm$ in the definition of 
 the particle creation operators $a^\dagger_\pm$.} 
 \beq
 \label{Projectors}
 \Ppm =  \frac{1 \pm \K}{2}.
 \eeq
 They satisfy the usual algebra of projection operators:
 \beq
 \label{Projectors2}
 \Pplus  + \Pminus =1, ~~~~~ \Ppm^2 = \Ppm  , ~~~~~  \Pplus\Pminus = 0.
 \eeq
 In addition,  one has 
 \beq
 \label{ProjectorsK}
 \K \Ppm = \Ppm \K = \pm \Ppm.
 \eeq
 Thus the Hilbert space decomposes as 
 \beq
 \label{Projectors3} 
 \CH = \Hplus  \oplus \Hminus , ~~~~~ | \psipm \rangle \in \Hpm , ~~~~~ \CK |\psipm \rangle = \pm |\psipm \rangle .
 \eeq
 One has 
 \beq
 \label{norms2}
 \brabra \psipm | \psipm \ketket = \langle \psi | \Ppm  \K \Ppm | \psi \rangle = \pm \langle \psipm | \psipm \rangle, 
 ~~~~~ 
  \brabra \psiminus  | \psiplus  \ketket  = \langle \psi | \Pplus \K \Pminus  | \psi \rangle =0 .
  \eeq

 If we project onto the sub Hilbert space $\Hplus$,     then all states $|\psi_n \rangle \in \Hplus$ have positive norm,  i.e. 
 $\K_n =+1$,   which implies positive probabilities $|c_n|^2 \geq 0$.        This positivity is preserved under time evolution:
 \beq
 \label{projectHilbert}
 \brabra \psipm (t) | \psipm (t) \ketket = \brabra \psipm (0) | \psipm (0) \ketket , ~~~~~
  \brabra \psiminus  (t) | \psiplus  (t) \ketket =  \brabra \psiminus (0) | \psiplus (0) \ketket = 0.
  \eeq
However this is insufficient to define a physically consistent quantum mechanics since $\K$ does not commute with $H$,  otherwise 
$H^\dagger = H$,  which is not the case by construction.       This implies that if one prepares a state at time $t=0$ that is in $\CH_{\plusdot}$,  then negative norm states
will be generated under time evolution:
\beq
\label{generationofneg}
e^{-i H t}  |\psi_{\plusdot} \rangle ~ \notin~ \CH_{\plusdot} 
\eeq
In order to have a consistent quantum mechanics one needs an additional {\it selection rule}  of sorts such that 
$e^{-i H t}  |\psi_{\plusdot} \rangle ~ \in~ \CH_{\plusdot}$.    At this point,  if such a selection rule exists,  it doesn't follow generally from pseudo-hermiticity,  but rather
must be model dependent.       In our model,   the free hamiltonian $H_0$ does commute with $\K$,  but the interacting part does not.     It is thus natural to consider this issue
in the interaction picture,  in particular in scattering theory,  which we turn to next.

  \def\bra{\langle}
  \def\ket{\rangle}

 \subsection{Specialization to our model:   Scattering theory and a  kinematic unitary regime}

 In the interaction picture  the hamiltonian is separated  into a free part $H_0$ and the interaction:
\beq
\label{Hint}
H = H_0 + H_{\rm int} .
\eeq
For the purposes of this section,   we again only consider a single component $\Phi$ as in Section IIA,   and consider the pseudo-hermitian  interaction
such as $H_{\rm int} = \int d^3 x \, (\Phi^\daggerk \Phi)^2$. 
Since $H_{\rm int}$ is local,   then far from the interaction region  the asymptotic  states in the far past and far future are 
eigenstates of the free hamiltonian,  and one can separate out this time evolution with $H_0$  which leads to a well-defined perturbation theory based on 
$H_{\rm int}$.      For hermitian quantum mechanics this leads to the standard formalism of scattering theory,   where positive probabilities lead to physically well-defined
and meaningful cross sections.

Let us adapt  this scattering formalism to pseudo-hermitian theories where the negative norm states cannot simply be projected out consistently.       
Define  the operator
\beq
\label{Omega}
\Omega (t) = e^{i H t} e^{-i H_0 t}  
\eeq
where as above  we have taken the reference time $t_0 =0$ for simplicity,  otherwise $t$ in the above equation should be replaced by $t-t_0$.          
Using $[ \K, H_0 ] = 0$,   one can easily show that 
\beq
\label{Omegadag} 
\Omega^\daggerk \Omega = 1.
\eeq
The so-called in and out states are defined by the M\o ller operators $\Omega_\pm$:
\beq
\label{inout}
|\psi \rangle_{\rm in} = \Omega_- |\psi \rangle, ~~~~|\psi \rangle_{\rm out } = \Omega_+ |\psi\rangle , ~~~~~\Omega_\pm = \lim_{t\to \pm \infty}  \Omega (t).
\eeq
Given  the $\K$-metric,   the S-matrix operator is now defined as follows:  
\beq
\label{Smatrix}  
_{\rm out} \brabra \psi' | \psi \ketket_{\rm in} \equiv  \langle \psi' | S | \psi \rangle  ~~ \Longrightarrow ~~  S = \Omega_+^\dagger \K \Omega_- = \K \, \Omega_+^\daggerk \Omega_- \, \, .
\eeq
Using $S^\daggerk = \Omega_-^\daggerk \Omega_+ \, \K $,  one finds 
\beq
\label{Sunitarity}
S^\daggerk S = 1.  
\eeq

The formula \eqref{Sunitarity} generalizes the usual statement of the unitarity of the S-matrix and  is equivalent to the fact that the norm of states is time independent,  i.e. \eqref{timeevolution}.   Thus by itself,  it does not resolve the  problem with negative norm states.   To address this,   one needs to consider physical quantities  that have a probabilistic meaning,  such as those in  the optical theorem.      To this end define the  usual $T$-operator  as 
\beq
\label{Ts}
S = \K + i T,   ~~~~~ S^\daggerk = \K - i T^\daggerk \,
\eeq
The $\K$ in the above equation comes from the $\K$ in the inner product if $T=0$.    The generalized unitarity relation \eqref{Sunitarity} expressed in terms of $T$ reads 
\beq
\label{optical}
i \( T^\daggerk  \K - \K T \) = T^\daggerk T
\eeq
Taking an inner product of the above $\langle \psi | \sim | \psi\rangle$ and inserting a complete set of states on the right hand side, $\K_\smallpsi$ cancels from both sides and one obtains  
\beq
\label{optical2} 
2  \, {\rm Im}  \( T_{\smallpsi \smallpsi} \) = \sum_{\psi'}   \K_{\smallpsi'} | T_{\smallpsi\smallpsi'} |^2 
\eeq
where we have defined  $T_{\smallpsi \smallpsi'} = \langle \psi | T | \psi'\rangle$.     The above is the generalization of the optical theorem for pseudo-hermitian theories.
If there are negative norm intermediate states $|\psi'\rangle$,  then they imply negative residues of kinematic poles in the S-matrix.   
It was first obtained by Cheng-Yang Lee \cite{CYLee} for scalar symplectic fermions \cite{LeClairNeubert},  where the precise nature of the pseudo-hermiticity is  essentially identical  to that of our bosonic model.     The above formula \eqref{optical2}  clearly shows the implications of  intermediate states with negative norm 
 $\K_{\smallpsi'} = -1$,  since for unitary theories all   $\K_{\smallpsi'} = +1$ and the right hand side is interpreted as the sum over positive transition amplitudes,  which leads to physically meaningful  cross sections.

In order to have an effectively unitary theory with positive probabilities,  an  additional selection rule is needed such that {\it only} intermediate states with 
$\K_{\smallpsi'} = +1$ contribute to the right hand of \eqref{optical2}.    There is no such selection rule  in  our model without adding some additional properties not
already contained in its definition.      However there exists such a selection rule based simply on kinematics,  as we now explain.      
 Our model is a quantum field theory with the multi-particle Hilbert space \eqref{states},  
thus as usual we factor out the overall energy-momentum conserving delta-functions and define the amplitudes $\CM$:
\beq
\label{withdeltas}
T_{\smallpsi\smallpsi'} = (2 \pi)^4  \delta^{(4)} \( p_\smallpsi - p_{\smallpsi'}\) \CM_{\smallpsi \smallpsi'} \, .
\eeq
Consider adding a mass term to $H$ such that the particles have mass $m$.  
Furthermore consider an initial state with only particles and  no anti-particles such that the in-state has positive norm,  and to further simplify matters let the in-state have only 
2 particles each of positive norm designated simply as $|++\rangle \equiv a^\dagger_+ a^\dagger_+ |0\rangle $.   Then the lowest energy intermediate state $|\psi'\rangle$ with negative norm 
that is consistent with the $U(1)$ charge $\qop$ conservation 
is $|+++- \rangle$ which requires particle/anti-particle  pair production $|+-\rangle$.   
We have used the fact that for a charge neutral operator $A$,   since $ [\qop, H] =0$,   one has 
\beq
\label{Qconservation}
 \brabra \psi' | A |\psi \ketket =0 ~~{\rm if} ~ \qop_{\smallpsi'} \neq \qop_\smallpsi .
 \eeq
If the total incoming energy is below the 4-particle threshold,  namely $s< (4m)^2$ where $s= (p_1 + p_2)^2$ is a  Mandelstam variable,   
then this process is kinematically forbidden.     Thus if one considers only 2 body scattering,   namely $2 \to 2$ particles,    then this process is unitary.   
The right hand side of \eqref{optical2} has only positive probability contributions which leads to standard formulas for the total cross section \cite{Peskin}:     
\beq
\label{optical3}
2 \, {\rm Im} \CM \( \kvec_1 , \kvec_2 \to \kvec_1 , \kvec_2 \) = 2 E_{\rm cm} p_{\rm cm} \, \sigma_{\rm tot} \( \kvec_1, \kvec_2 \to {\rm anything} \).
\eeq

In summary,  for 2-body elastic scattering of particles,  if the total energy is below the threshold for pair production,  the theory is effectively unitary.  
This result is analogous to the  low energy scattering of electrons $e_-$ below the threshold for $e_+ e_-$ pair production,  commonly referred to 
as  ``electron–electron M\o ller  elastic scattering"  $e_- e_- \to e_- e_-$, 
which is known to be unitary if one ignores production of  electron/positron pairs.\footnote{In full quantum electro-dynamics (QED) 2 body scattering can also radiate  photons,
but this amounts to almost negligible effects.   In any case our model consists of only particles and anti-particles since we have not gauged the U(1).} 
 There is a  difference between  our model and QED  since  both  electrons and positrons  have positive norm,  nevertheless the kinematic constraint still holds
 and the formula \eqref{optical} applies to both.   In the non-relativistic limit where all scattering is elastic,   namely the number of incoming and outgoing particles is the same, 
 our model is unitary.

A similar,  but essentially different,  selection rule that restores unitarity is to consider the model at finite density by introducing a chemical potential $\mu$:
\beq
\label{chemicalpotential}
H \to H - \mu \qop = H-\mu \(N_+ - N_- \).
\eeq
For positive $\mu$,   it costs extra energy to add an anti-particle with $N_- = \Nbar > 0$ whereas adding particles $N_+$ is energetically cheap.   Thus with high enough
chemical potential,  production  of anti-particles  with negative norm is suppressed.

The above conclusions open up the possibility of applying the  pseudo-hermitian bosonic models of this paper,   and also symplectic fermions \cite{LeClairNeubert},  to 
condensed matter physics,    where energies are low enough,  in fact typically non-relativistic,   which are effectively unitary since one can ignore pair production at these 
energies.   Symplectic fermions have a Fermi surface,  thus the above discussion of $\CC$-conjugation corresponds to particle-hole symmetry.     
Some such applications were discussed for symplectic fermions in \cite{LeClairNeubert}.    Let us also mention that 
since the groups Sp(4) = SO(5),    the N=2 component symplectic fermion has order parameters for anti-ferromagnetism,   superconductivity,  and also a potential pseudo-gap,
and can be studied as a toy model of high $T_c$ superconductivity \cite{KapitLeClair}  where the role of SO(5) was first proposed by S.-C. Zhang \cite{SCZhang}.\footnote{ In light of the above remarks,  the  main shortcoming  of the work \cite{KapitLeClair}   is not  necessarily  that it is non-unitary,   but rather  that it is formulated in the continuum without a lattice,  whereas the lattice  is known to play a fundamental role in obtaining a d-wave  superconducting gap.   As a toy model of superconductivity,  it still has a Fermi surface,  and  it is distinguished by the fact the 4-fermion interactions closely parallel the interactions in the Hubbard model,  and since they are marginal,    they imply a fermionic critical point in 2+1 dimensions that can be reached
in $4-\epsilon$ dimensions.   This critical point was studied to 2-loops in \cite{LeClairNeubert}.}   For more  recent applications of symplectic fermions to condensed matter 
see \cite{Gao}.

\section{Higher order RG beta-functions based on the Operator Algebra (OPE)}

In this Section we show how to compute the RG beta-functions based solely on the OPE in \eqref{JOPE}.   
For our model,  the parameter $\kappa = 1$,  however we display it un-evaluated since,   as we will see,   powers of $\kappa$ counts loops.   
The definition of our model in Section II very closely resembles current-current perturbations  in 2D based on a Lie algebra,    
 in fact was modeled after it.   We review  this  2D case in the Appendix.    
The role of the independent  left/right moving currents $J^a , \bar{J}^b$ in 2D  is played by the independent operators  $J^a ,  \Jtilde^b$. 
The OPE \eqref{JOPE} also closely parallels the 2D OPE in  \eqref{CurrentOPE},   thus the Lie-algebraic structure of higher orders in perturbation theory
is identical to the 2D case.   This implies that we can organize higher orders in perturbation theory using the same ``OPE diagrams"  introduced in \cite{Gerganov}.
The important difference  between our 4D model and 2D current-current perturbations are the integrals involved in perturbation theory.    
Thus we work out these integrals up to 3-loops in order to compare with the 2D beta functions based on the same Lie algebra,  in this case SU(2). 
As we will show,  up to 3 loops the beta functions completely agree in 4D and 2D.    Since the Lie-algebraic structure is identical,   this will allow us to 
propose an all orders beta function based on the results in \cite{Gerganov}.

The SU(2) symmetric case corresponds to 
$g_1 = g_2 =  g_3=g $ since then the perturbation $g\, \sum_ a  J^a \Jtilde^a$ is built on the quadratic Casimir.      
We are more interested in the case where SU(2) is broken to U(1),  where $g_1 = g_2 \neq g_3$ considered below.  
We will first  compute the beta functions to 3-loops for the fully anisotropic case $g_1 \neq g_2 \neq g_3$,  
since it's easier to work out the beta functions for this  case first and then set $g_1 = g_2$.      
Renormalization group beta-functions are generally prescription, or scheme, dependent.     However for marginal perturbations,   the  one and two loop contributions are universal.\footnote{To see this,   consider a single coupling $g$ with 
$\beta_g = b_2 g^2 + b_3 g^3 + \ldots$ where $b_{2,3}$ are coefficients.      The prescription dependence corresponds to a redefinition of the coupling $g' = g'(g)$.    Let $g' = g + c g^2 + \ldots$.   One easily sees that 
$\beta' (g') = b_2 g'^2 + b_3 g'^3 + \ldots.$}   
The implicit renormalization prescription of this article  is based on the OPE \eqref{JOPE},  and doesn't explicitly  rely on any epsilon-expansion, nor Feynman diagrams for scalar fields,  but it should still be viewed as a specific prescription,  and other prescriptions could lead to different expressions beyond 2 loops.\footnote{We do not address whether this OPE prescription can be 
shown to be equivalent to 
Feynman diagram perturbation theory and dimensional regularization to all orders in perturbation theory.       However to 1-loop it was shown that these two-prescriptions do indeed agree  for the closely related symplectic fermions \cite{LeClairNeubert}.       Referring to the title of Wilson-Fisher's paper \cite{WilsonFisher},   we work in precisely $4$ dimensions,  not $3.99$!}    The significant properties such as the existence of fixed points,  flows between them,  and limit-cycle behavior are expected to be independent of the particular renormalization scheme. 

We first review the 1-loop result in \cite{ALDolls},   then describe how to extend this to higher orders.     The starting point is the perturbative expansion about the conformal field theory for
$\langle \X \rangle$ where $\X$ is an arbitrary operator (it can in fact to be taken to be the identity):  
\beq
\label{XallOrders}
\brabra \X \ketket   = \sum_{n=0}^\infty  \frac{(- 2 \pi^2)^n}{n!}  \sum_{a_i} g_{a_1} g_{a_2} \cdots g_{a_n} \,
\int d^4 x_1 d^4 x_2 \cdots d^4 x_n \, \brabra  \CO^{a_1} (x_1)  \CO^{a_2} (x_2)  \cdots \CO^{a_n} (x_n)  \, \X \ketket_0 ,
\eeq
where $\brabra X \ketket _0$ is the unperturbed CFT result.   
The $n$-th  term  gives a contribution to the $(n-1)$-loop  beta  function.

\subsection{1-loop}

For the one-loop beta function one only needs the OPE's of the operators $\CO^a$ themselves.        
Using \eqref{JOPE} one finds 
\beq
\label{OOPE}
\CO^a (x) \CO^b (y) =   \frac{\delta^{ab} }{|x-y|^8} \,\, \frac{\kappa^2}{64 \pi^8}  -   \inv{16 \pi^4 |x-y|^4 }  \,\,  C^{ab}_c  \, \,\CO^c (0) + \ldots 
\eeq
where there is no sum over $c$ for SU(2), we used $\delta^{ab} f^{abc} = 0$,  and the only non-zero terms are 
\beq
\label{Cabc}
C^{ab}_c =  (f^{abc})^2 = 4   ~~~~~ \forall ~ a \neq b \neq c .
\eeq
To order $g^2$: 
 \beq\label{ope.4} 
   \brabra  \X\ketket  = \brabra  \X\ketket _0
   - 2\pi^2 \sum_c g_c \int d^4x\,\brabra  \CO^c (x) \, \X\ketket_0 
   + 2 \pi^4   \sum_{a,b} g_a g_b  \int d^4x \int d^4y\,
   \brabra  \CO^a (x)\,\CO^b(y)\,\X  \ketket_0 + \dots \,, 
\eeq
Using the OPE \eqref{OOPE} 
along with 
\beq 
\label{logInt} 
\int_a  \frac{d^4x}{x^4}=-2\pi^2\log a, 
\eeq
where $a$ 
is an ultraviolet cut-off, one finds
\beq\label{ope.5}
   \brabra \X\ketket 
   = \brabra \X\ketket_0 
   - 2\pi^2 \sum_{a,b,c}  \( g_c -\inv{8} C^{ab}_{c} g_a g_b \, \log a \) \int d^4 x\,\brabra \CO^c (x)\, \X \ketket_0
   + \dots \,. 
\eeq
The ultraviolet divergence is removed by letting 
$g_c \to g_c + \tfrac{1}{8} C^{ab}_c  g_a g_b \, \log a$.   This leads to 
\beq
\label{betaA}
\beta_{g_a} \equiv \frac{dg_a}{d \ell}   = \inv{8} \sum_{b,c} \, C^{bc}_a\, g_b g_c 
\eeq
where increasing $\ell = \log a $ is the flow to low energies.   
Thus to 1-loop we have 
\beq
\label{1loopBetas}
\beta_{g_1} = g_2 g_3 ,
~~~ 
\beta_{g_2} = g_1 g_3 ,
~~~ 
\beta_{g_3} = g_1 g_2 .
\eeq
The above calculation indicates that our model is renormalizable to 1-loop,   meaning that no additional operators,  
such as $J^a J^a$ or $\Jtilde^a \Jtilde^a$,  are generated since the OPE \eqref{OOPE} closes.  

Below we will be mainly interested in the case $g_1 = g_2 \neq g_3$ which breaks SU(2) to U(1),   however let us comment on the fully anisotropic case to 1-loop and its 
integration since this  provides insights on  the main properties of these flows based on comparison with known dynamical systems,  in this case rigid body motion.         
The above equations \eqref{1loopBetas} can be explicitly integrated as a function of the scale $\ell$ by using certain RG invariants $Q$ satisfying 
\beq
\label{Qinv}
\sum_g \beta_g \d_g Q = 0 .
\eeq
There are 3 such invariants,   only two of which are linearly independent:
\beq
\label{Qs}
Q_1 = g_2^2 - g_3^2, ~~~~~ Q_2 = g_3^2 - g_1^2,~~~~~~Q_3 = g_1^2 - g_2^2,  ~~~~~~~~~Q_1 + Q_2 + Q_3 =0.
\eeq
By substituting these into $\beta_{g_3} = dg_3 /d\ell$ to eliminate $g_1, g_2$ in favor of $Q$'s,   the  solution $g_3 (\ell)$ of the resulting differential equation
can be expressed in terms of the Jacobi elliptic functions {\bf{ns}}  and {\bf{cs}}   which are {\it doubly}  periodic,  where the two periods depend the $Q$'s.  
 (See \cite{Elliptic} for explicit formulas for $g_3 (\ell, Q)$.)   
The appearance of elliptic functions could have been anticipated from the observation that if $\ell$ is treated as a time and the beta functions 
\eqref{1loopBetas} viewed as dynamical equations with $\dot{g} \equiv  dg/d\ell = \beta_g$,   then they can be mapped to rigid body dynamics in the absence of torque, which was solved by Jacobi.   The integrals of motion for this dynamical system are the RG invariants $Q$.  
To see this,  let $\vec{L} = (L_1, L_2 , L_3)$ denote the 3 components of the classical angular momentum and $\dot{L}$ their  time derivatives.    The Euler equations are 
\beq
\label{Ldot}
\dot{L}_1 = \( \inv{I_3} - \inv{I_2} \) L_2 L_3,  ~~~ \dot{L}_2 = \( \inv{I_1} - \inv{I_3} \) L_3 L_1, ~~~~\dot{L}_3 = \( \inv{I_2} - \inv{I_1} \) L_1 L_2 ,
\eeq
where $I_a$ are principal moments of inertia.  
    Identifying the moments $I_a$ as follows,   together  with a rescaling of $L_1, L_2$:
\beq
\label{Iident}
I_2 = \frac{I_1}{1- 2 I_1}, ~~~I_3 = \frac{I_1}{1-I_1}, ~~~~~~~~~~  L_{1,2} \to  \frac{i}{\sqrt{2}} \,  L_{1,2} , 
\eeq
one obtains the 1-loop RG flow equations \eqref{1loopBetas} for $g_a$  from \eqref{Ldot}  with  $L_a  \to  g_a$.    Let us also point out that 
 a quantum spin version of  this is the anisotropic Heisenberg type hamiltonian:
\beq
\label{Hspin}
H = \frac{L_1^2}{2 I_1} + \frac{L_2^2}{2 I_2} +  \frac{L_3^2}{2 I_3} ,
\eeq
which has some resemblance to the marginal perturbations in our model of QFT as defined in \eqref{perturbations}. 
The quantum equations of motion,   $\dot{L_a} = i [H,L_a]/\hbar$,  
leads to the equations \eqref{Ldot}  in the classical  limit if one assumes the $\vec{L}$ satisfy the usual quantum angular momentum algebra.

\subsection{2-loops} 

\begin{figure}[t]
\centering\includegraphics[width=.6\textwidth]{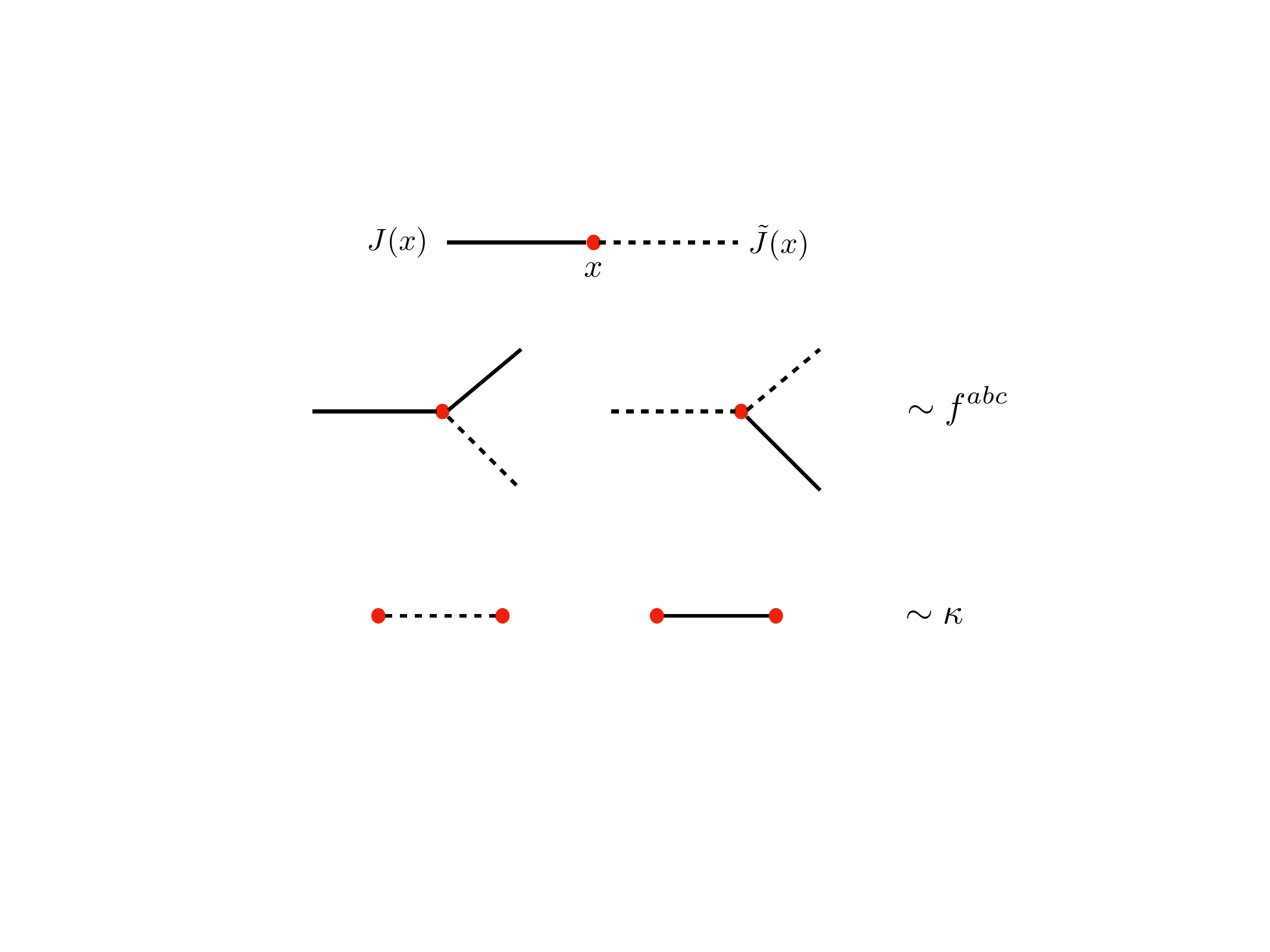}
\caption{OPE diagram building blocks.   Each red dot signifies a spacetime point $x$.    Solid lines refer to $J$ whereas 
$\tilde{J}$ is represented by a dotted line.}
 \label{diagrams}
\end{figure}

\begin{figure}[t]
\centering\includegraphics[width=.45\textwidth]{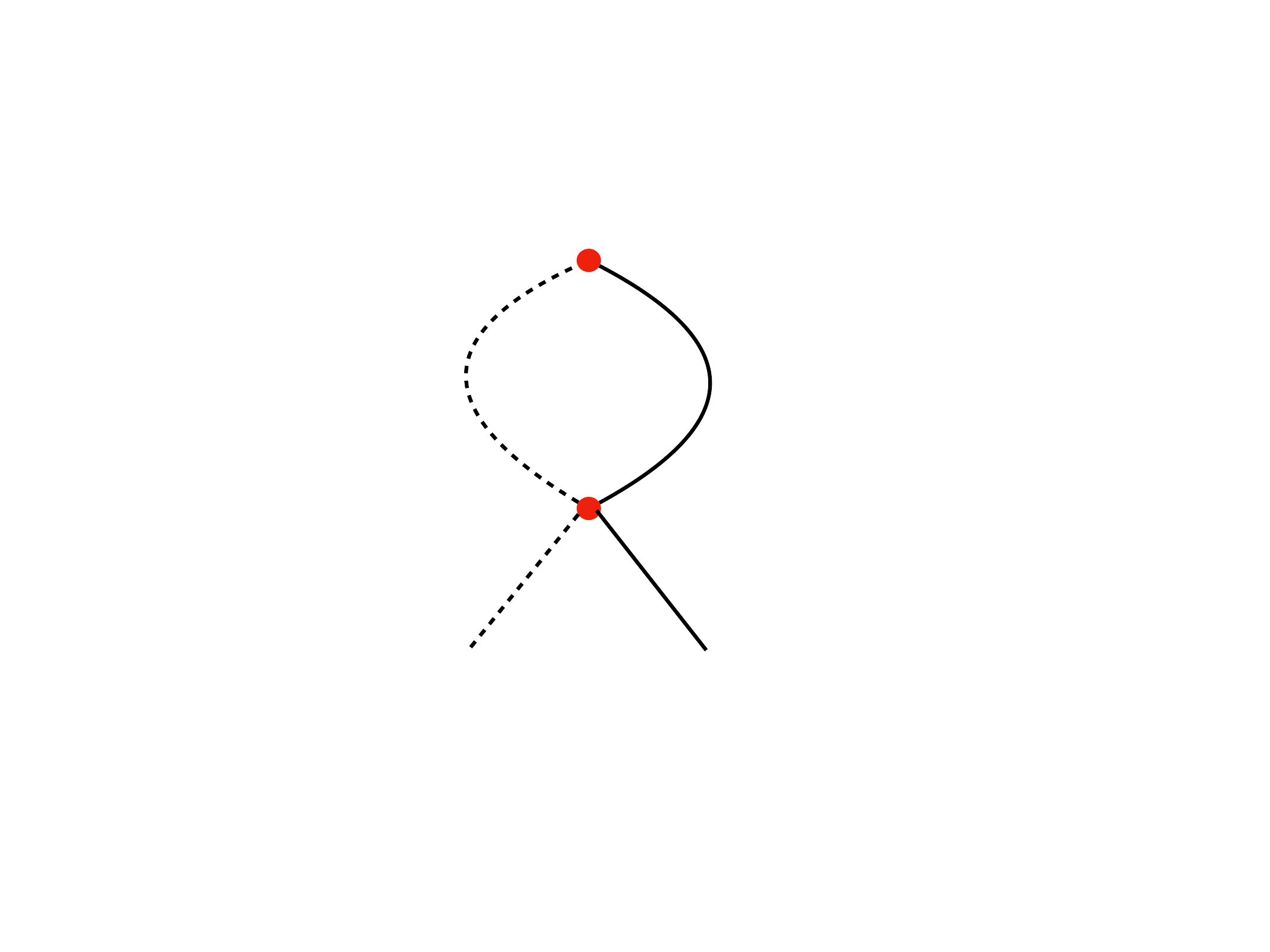}
\caption{1-loop diagram}
 \label{1loop}
\end{figure}

As we did above at 1-loop,    at higher orders we use the OPE prescription to isolate $\log a$ divergences that contribute to the beta-functions.     There are also $(\log a)^n$ divergences for $n>1$,   which we interpret as arising from lower orders in perturbation theory which are already accounted for in the  lower order beta-function.      It turns out that the OPE \eqref{OOPE} for the operators $\CO^a$   is not sufficient to extract the appropriate $\log a$ contributions  for 2-loops and beyond:    one  must split $\CO^a$ into the product of $J^a$ and $\Jtilde^a$ and use their OPE's separately.        It is useful to introduce diagrams that indicate the OPE structure that leads to a $\log a$ divergence of interest,   with the building blocks shown in 
 Figure \ref{diagrams}.    The Lie-algebraic structure of these diagrams are identical to the OPE diagrams for  2D current-current perturbations introduced in \cite{Gerganov}.\footnote{Unlike Feynman diagrams,   these diagrams are not a prescription for calculating the full correlation function,   but are useful for isolating $\log a$ divergences.}     The 1-loop contribution is shown in Figure \ref{1loop}.

\begin{figure}[t]
\centering\includegraphics[width=.4\textwidth]{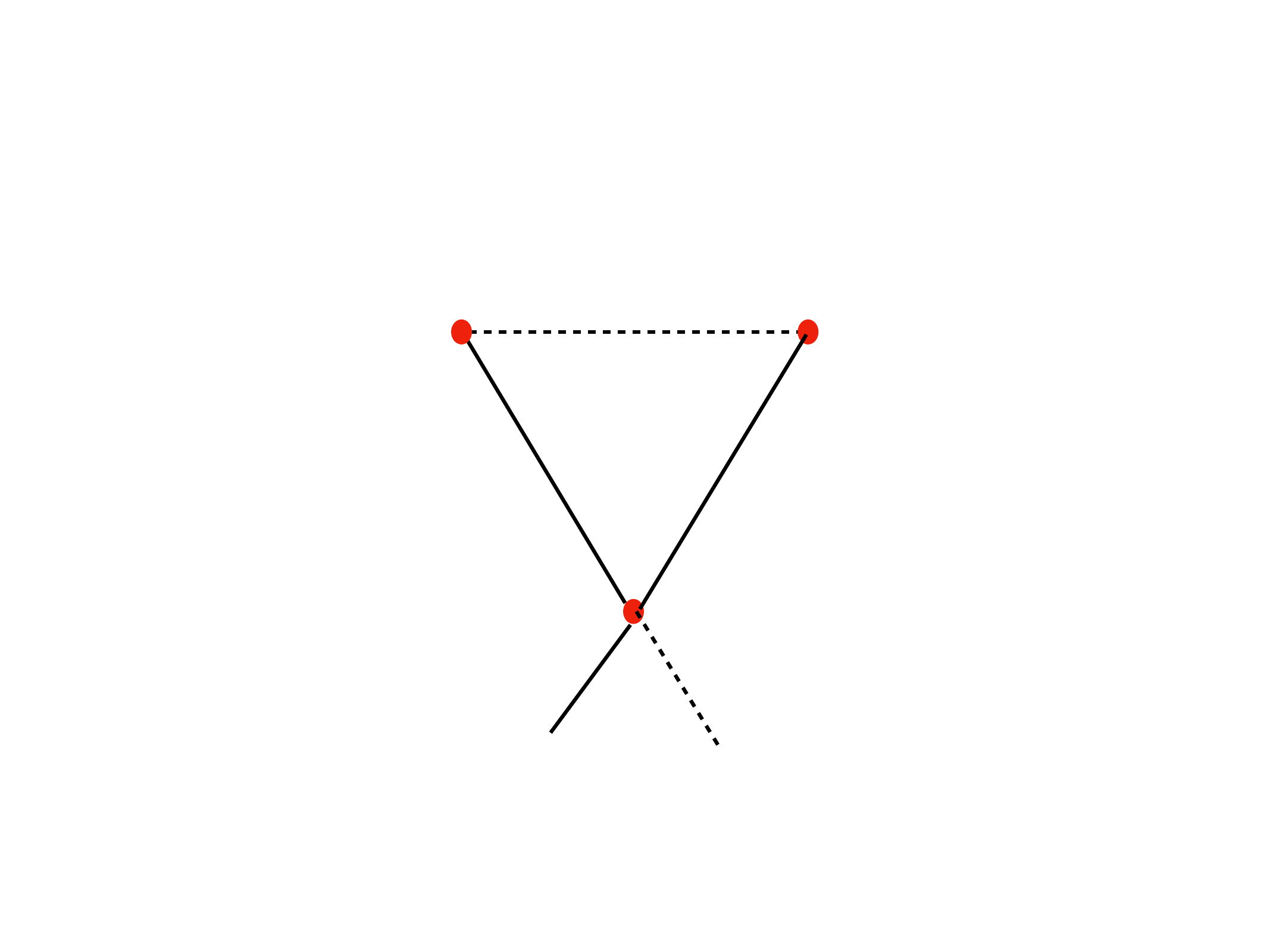}
\caption{The only 2-loop diagram that gives an additional contribution to the beta-functions.}
 \label{2loop}
\end{figure}

To order $g^3$ one has 
\beq
\label{2loop1}
\brabra  \X \ketket_{{\rm 2-loop~ term}} = - \frac{ (2 \pi^2)^3 }{3!} \sum_{a,b,c} g_a g_b g_c \,  \CI^{abc}   
\eeq
where 
\beq
\label{2loop2}  
\CI^{abc}  =  \int  d^4x_1\, d^4x_2\, d^4x_3\,  \brabra  \CO^{a}(x_1) \CO^b (x_2)  \CO^c (x_3)  \X \ketket.
\eeq
Consider the contribution where $\CO^c (x_3)$ is left over after OPE's:    
\beq
\label{2loop3}
\CI^{abc}  \sim  - \frac{ \kappa}{128 \pi^8 }\,  \delta^{ab} \sum_e f^{ace} f^{bce} 
\int d^4x_1 \,d^4x_2 \, d^4x_3 \, 
\inv{ |x_1 - x_2|^4 |x_1 - x_3|^2 |x_2 - x_3|^2 } \, \brabra  \CO^c (x_3) \X \ketket_0. 
\eeq
This corresponds to the OPE diagram in Figure \ref{2loop}.   
One can perform  the integral over $x_1, x_2$ by shifting $x_1 \to x_1 + x_3$,    $x_2 \to x_2 + x_3$,  and using the integral:
\beq
\label{2loop4} 
\int d^4x_2  \, \inv{|x_1 -x_2 |^4 |x_2 |^2} = \frac{\pi^2}{2} \inv{|x_1|^2}.
\eeq
Then integrating over $x_1$ using \eqref{logInt} gives a $\log a$ divergence.        This contribution comes with a combinatorial factor of $2 \times 3!$,  where the extra $2$ comes from the additional diagram that exchanges $J$ with $\Jtilde$.  As for 1-loop we absorb this divergence in the couplings $g_a$.   
This leads to the beta functions 
\beq
\label{2loop6} 
\beta_{g_1} = g_2 g_3 - \frac{\kappa}{4} \, g_1 (g_2^2 + g_3^2 ) , ~~~~{\rm plus ~ cyclic ~ permutations~ of ~ 1,2,3}.
\eeq

\begin{figure}[t]
\centering\includegraphics[width=.5\textwidth]{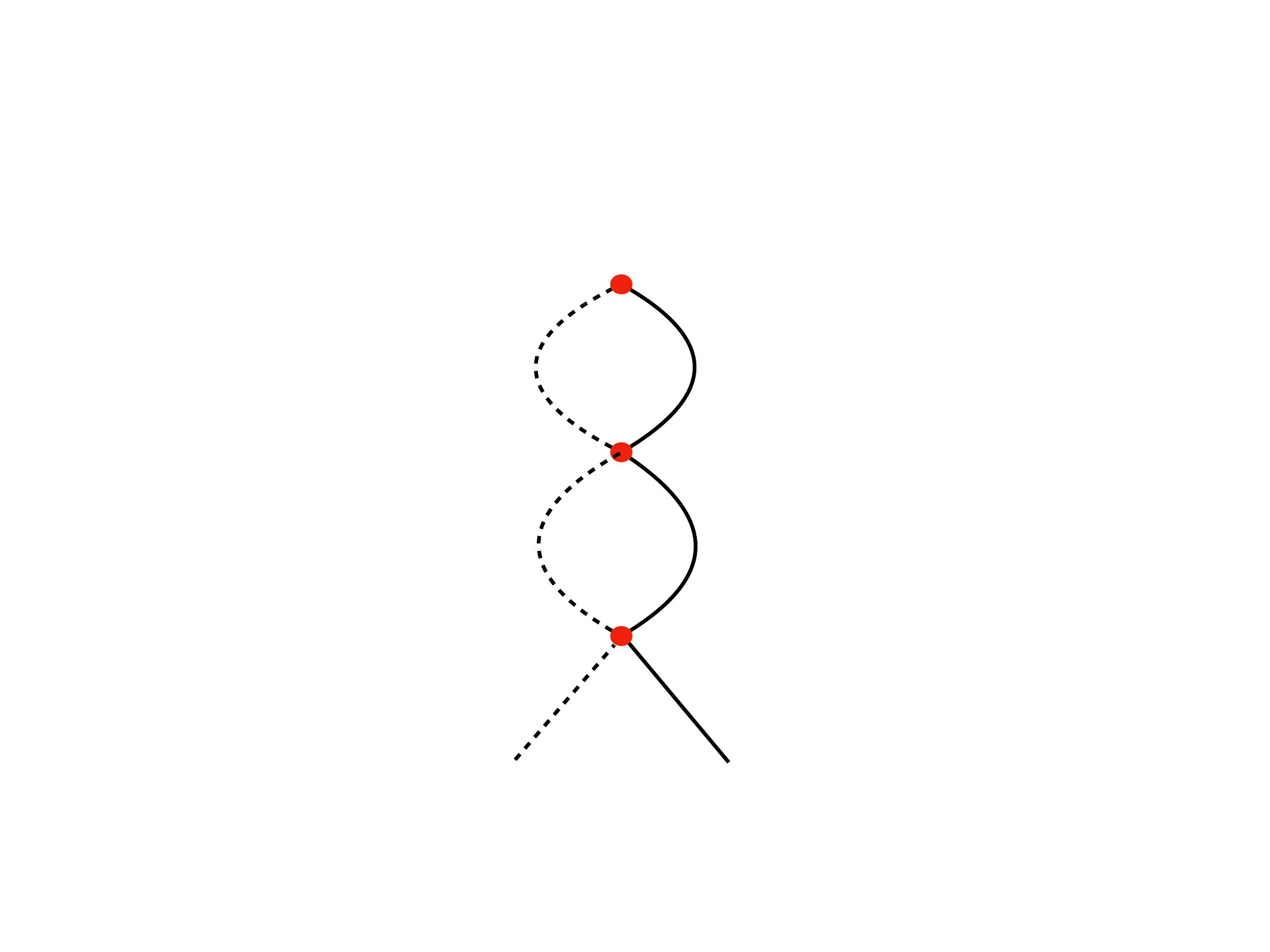}
\caption{2-loop diagram  where the divergence is the product of two 1-loop divergences.}
 \label{double1loop}
\end{figure}

\begin{figure}[t]
\centering\includegraphics[width=.5\textwidth]{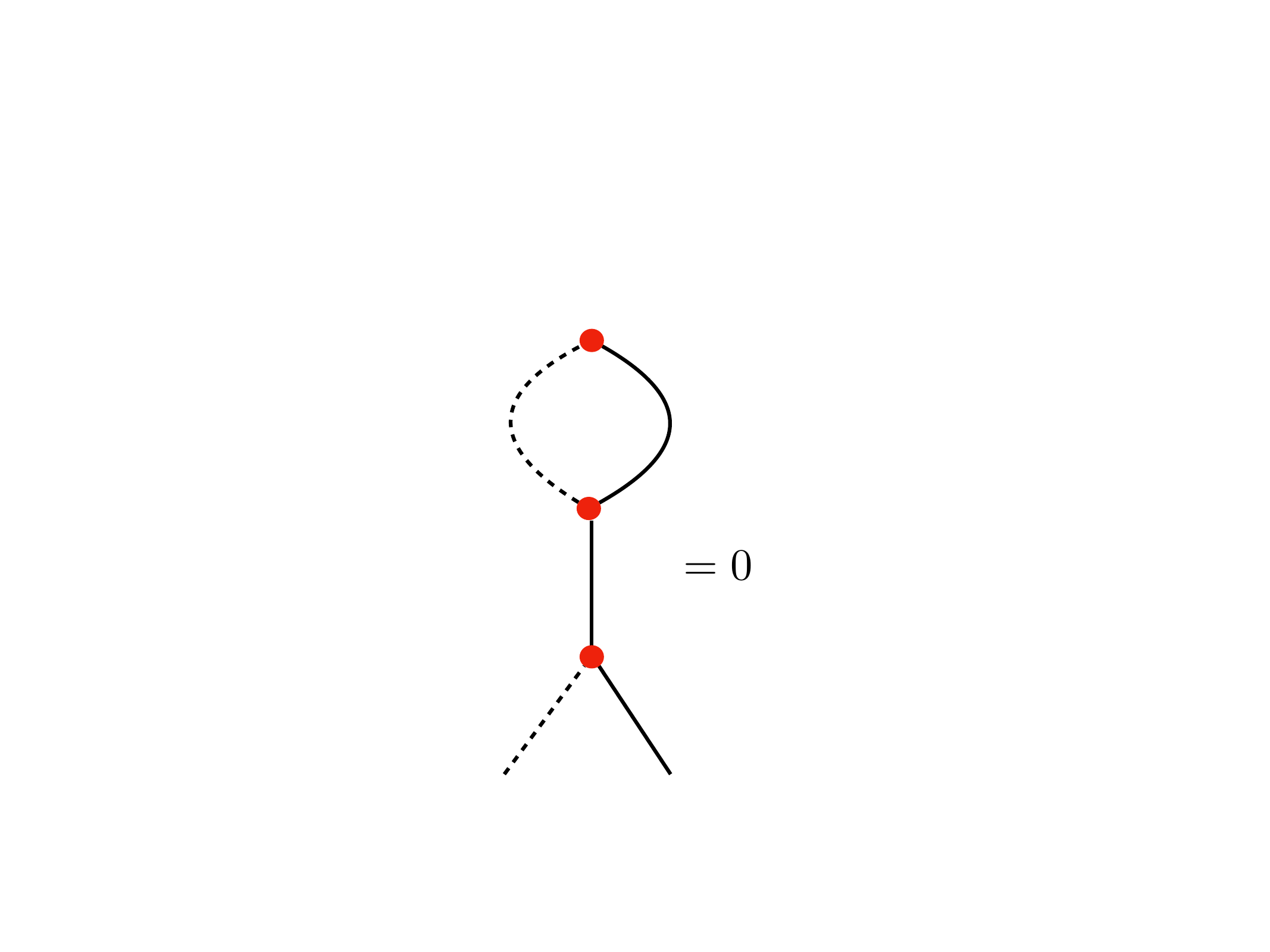}
\caption{2-loop diagram that gives zero contribution to the beta-function since it is proportional to $\delta^{ab} f^{abc}$.}
 \label{Type2}
\end{figure}

\begin{figure}[t]
\centering\includegraphics[width=.4\textwidth]{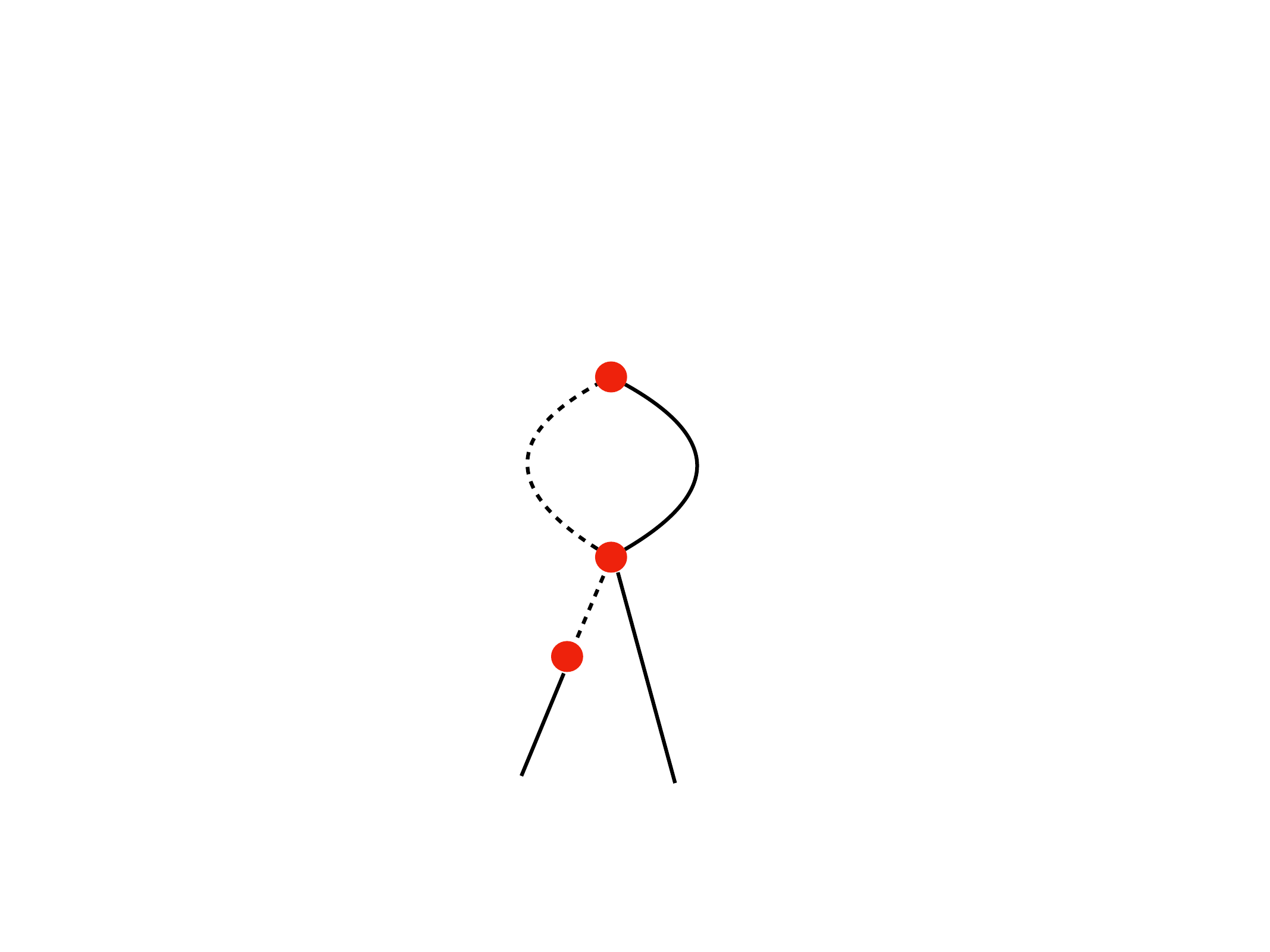}
\caption{2-loop diagram that potentially gives rise to  additional $JJ$ terms in the action.}
 \label{JJ}
\end{figure}

The above result  \eqref{2loop6},   based on Figure \ref{2loop},   is the only contribution at order $g^3$.      For instance the diagram in 
Figure \ref{double1loop} gives a $\log^2 a$ divergence which is already incorporated in the 1-loop beta function.  
The diagram in Figure \ref{Type2} gives zero since it is proportional to $\delta^{ab} f^{abc}$.        Figure \ref{JJ} diagram could potentially lead to new terms  $J^a (x) J^a (x)$ in the action rendering our original model non-renormalizable,  however as is evident from the diagram,  the two $J$'s  are at different spacetime points,   and furthermore leads to $\log^2 a$ divergences.

\subsection{3 loops} 

\begin{figure}[t]
\centering\includegraphics[width=.6\textwidth]{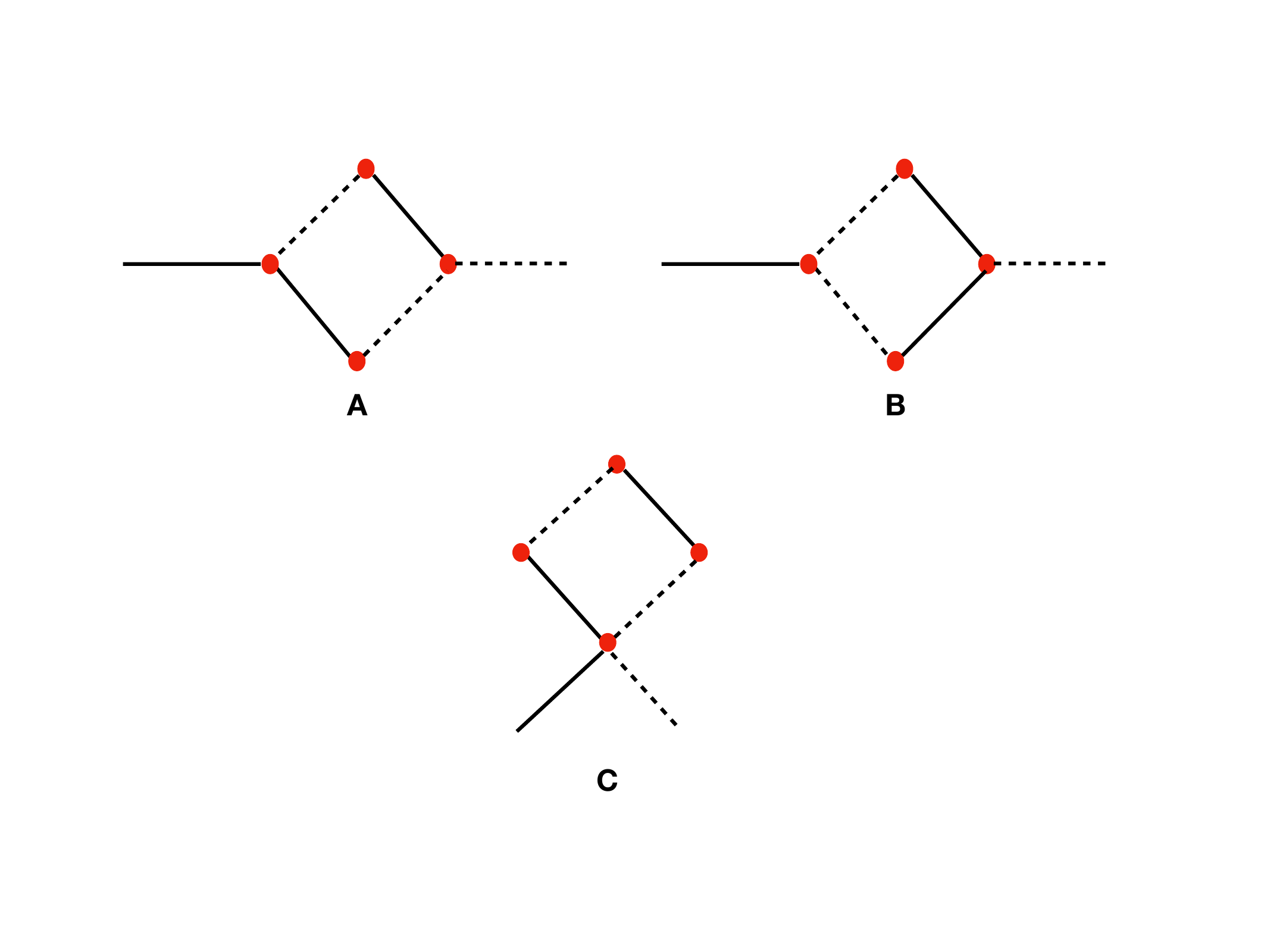}
\caption{3-loop diagrams of  type A, B and C.}
 \label{3loop}
\end{figure}

There are 3  OPE diagrams that contribute to order $g^4$,   shown in Figure \ref{3loop}.  
For each diagram the combinatorial factor is $2 \cdot 4!$.   The type A and C diagram contributions can be evaluated using the same integrals \eqref{logInt} and \eqref{2loop4} as for 2-loops.     This gives 
\beq
\label{3loopAC}
\beta_{g_a}^{\rm 3-loop\, A} = \beta_{g_a}^{\rm 3-loop\, C}
= \frac{\kappa^2}{32} \sum_{b\neq c \neq a}  g_b^3 g_c ~ .
\eeq

The type B diagram is different since the OPE's involving $f^{abc}$ are internal to the diagram.    In this case one needs 
\beq
\label{integrals}
\int \frac{d^4 y}{|x-y|^4 |y|^4} =  \frac{\pi^2}{|x|^4}, ~~~~~
\int  \frac{d^4 y} {|x-y|^2 |y|^4} =  \frac{\pi^2}{4|x|^2} .
\eeq
This leads to 
\beq
\label{3loopB}
\beta_{g_a}^{\rm 3-loop\,B}
= \frac{\kappa^2}{32} \sum_{b\neq c \neq a}   g_a^2 g_b g_c ~ .
\eeq
Putting this all together,   to 3-loops the beta functions are 
\beq
\label{beta1All}
\beta_{g_1} =  g_2 g_3  - \frac{\kappa}{4} g_1 \( g_2^2 + g_3^2 \)  + \frac{\kappa^2}{16} \( g_1^2 g_2 g_3 + g_2^3 g_3 + g_3^3 g_2 \) 
\eeq
plus cyclic permutations of $1,2,3$ for $\beta_{g_2}$ etc.

\vfill\eject

\def\k{\kappa}

\subsection{SU(2) broken to U(1) and higher orders}

Now we consider the case where the marginal perturbations break SU(2) to U(1).      Comparing the 3-loop result to the 2D current-current models  
in \cite{Gerganov} will
allow us to propose a beta function to all orders.
  
Consider first the fully isotropic case $g_a = g ~ \forall a$.      Based on \eqref{beta1All} the beta function to three loops  is 
\beq
\label{isotropic}
\beta_g = g^2 - \frac{\kappa}{2} g^3 + \frac{3 \kappa ^2}{16} g^4 + \ldots 
\eeq
The only  fixed point is at $g=0$  which is marginally relevant or irrelevant depending on the sign of $g$.  
Recalling that our conventions for the RG flow are  such that increasing $\ell$ corresponds to a  flow to low energies,   when 
$g>0$  one flows to $g=0$ in the UV where the theory is the asymptotically free theory of the $\Phi,  \Phitilde$ fields.
More importantly,   one easily recognizes the first terms of a geometric series in \eqref{isotropic}:
\beq
\label{betasym}
\beta_g = g^2 - \frac{\kappa}{2} g^3  + \frac{3 \kappa^2}{16}  g^4 + \ldots  = 
\frac{g^2}{(1 + g \kappa/4)^2} .
\eeq 
   The above all-orders beta function is identical to the result in \cite{Gerganov} for fully isotropic 
SU(2).     This indicates that the OPE diagrams of the last section have nested geometric series which allows for  their resummation,   even though they are not as transparent 
as for the isotropic case in \eqref{betasym}.

Now we turn to  $\SU2$ symmetry broken to U(1).        
Setting $g_1 = g_2 \neq g_3$  in \eqref{beta1All},  to 3-loops  one has 
\barray
\label{2loop7} 
\beta_{g_1} &=& g_1 g_3 - \frac{\kappa}{4} g_1 (g_1^2 + g_3^2 ) + \frac{\kappa^2 }{16} \, g_1^2 (g_3^3 + 2 g_3 g_1^2 ) + \ldots 
\\
\beta_{g_3} &=&  g_1^2 - \frac{\kappa}{2}\,  \, g_1^2 g_3  + \frac{\kappa^2}{16} \, g_1^2  ( 2 g_1^2 + g_3^2 ) + \ldots
\earray
For describing the RG trajectories in the next section,    the analysis is greatly simplified by an  RG invariant $Q(g_1, g_3 )$,  since this does not require explicitly  integrating the beta functions as a function of scale $\ell$,   since RG trajectories are constant $Q$ contours.      Such an RG invariant $Q$ is defined as satisfying \eqref{Qinv},   and when 
$g_1 = g_2$ there is only one such invariant as can be seen from the 1-loop result \eqref{Qs}.      
To 1-loop let us define  this single  RG invariant as $Q = g_1^2 - g_3^2$.       
This invariant is not spoiled by the above higher loop computations,   however it does receive higher loop corrections:
\beq
\label{Q3loop}
Q = \( g_1^2 - g_3^2 \) \(1 + \frac{\kappa}{2} g_3 + \frac{\kappa^2}{16} (3 g_3^2 + g_1^2 )  + \ldots \).
\eeq
Here \eqref{Qinv} is satisfied to order $g^5$,   which implies it is valid to order $g^4$:
\beq
\label{Q3loopInv}
\sum_{g=g_1, g_3} \,  \beta_{g}  \d_g Q = \CO(g^5) .
\eeq

Based on the above results,   we can  now propose formulas for the beta functions to all orders in the coupling.   
This is possible based on the following reasoning.   (i)  The Lie-algebraic structure of the OPE diagrams in 4D and 2D is identical,   in fact by construction. 
The general formulas in \cite{Gerganov}  allow one to deal with arbitrary anisotropy in the couplings.    
(ii)  We explicitly checked that the integrals in 4D work out to give the same beta function as in \cite{Gerganov} for 2D current-current perturbations  up to 3 loops.  
  This aspect came as a surprise,   and we don't have a separate argument for why this is correct  besides 
the explicit 3-loop calculations we performed.     Some simple checks indicate that  this structure persists to higher orders,   however we did not carry out these calculations 
explicitly beyond 3 loops.       A further constraint,  or consistency check,   is that we expect the RG invariant $Q$ to persist to higher orders.      We can thus borrow the result from \cite{Gerganov}:
 \beq
\label{betaAllorders}
\beta_{g_1} = \frac{g_1 (g_3 -g_1^2 \k/4)}{(1-\k^2 g_1^2/16)(1+ \k g_3/4)}, ~~~~~
\beta_{g_3} = \frac{g_1^2 (1-\k g_3/4 )^2}{(1 - \k^2 g_1^2/16)^2 }.
\eeq
One can easily check that to order $g^4$,   the above agrees exactly with the 3-loop results \eqref{2loop7}.   
Also,   when $g_1 = g_3$ one recovers the fully isotropic case  \eqref{betasym}.

What is not so obvious,  and rather non-trivial,    is that these beta functions preserve the RG invariant $Q$,  which to all orders is now:
\beq
 \label{QAllorders}
 Q = \frac{g_1^2 - g_3^2}{(1- \k g_3/4)^2 (1- g_1^2 \k^2/16)} .
 \eeq
 The invariance condition  \eqref{Qinv} is {\it exactly} satisfied to all orders for this non-perturbative $Q$.\footnote{  
 For the fully anisotropic case $g_1 \neq g_2 \neq g_3$ we could not  readily  find such  RG invariants  to all orders that extend  the 1-loop results \eqref{Qs},   however nor could we show that they don't exist.}    The above beta functions do not constitute a gradient flow  
 where $\beta_{g_i} = - \d_{g_i}  \CH (g)$  for some height function $\CH(g)$,   since $\d_{g_1} \beta_{g_3} \neq \d_{g_3} \beta_{g_1}$.     This  is consistent with the existence of cyclic RG flows we find below since gradient flow is often associated with flows that are consistent with various forms of c-theorems.   
The above  beta functions and $Q$ have poles at the points $g_{1,3} = \pm 4$.    These are self-dual points under the strong-weak coupling duality described below, 
 equation \eqref{duality1},   and the RG flows pass through them smoothly,  as we will show below.      
 
The remainder of this article is based on the all-orders formulas  in equations  \eqref{betaAllorders},\eqref{QAllorders}.     
 Above,  we provided strong arguments for the validity of these formulas.   Although we did not write out a complete proof,   since the 4D integrals involved worked out up to 
 3 loops,  and higher orders involve essentially the same kinds of integrals,   a complete proof can be formulated along the lines in \cite{Gerganov}.  
   In 2D,   there exists non-perturbative checks of the  above beta functions \eqref{betaAllorders} 
 based on known massless  flows between CFT's which have  an exact Bethe-Ansatz solution reviewed in the Appendix.  The latter check is based on the correct 
 relation between the anomalous scaling dimension $\Gamma_\UV$ of the relevant perturbation of the CFT in the UV and the dimension $\Gamma_\IR$ of the operator by which it arrives to the IR fixed point CFT.      This relation will be generalized for our 4D model in the next Section.       The only way we can see that the formulas 
\eqref{betaAllorders} 
 are not completely correct  is if there  are  additional contributions to the beta functions in a different prescription which  are  missed by our prescription based on the OPE.       
At worse,    the above beta functions represent a re-summation of important contributions to the beta functions and can serve to provide some non-trivial checks of the RG flows we propose below.      For the above reasons,   one should perhaps still view \eqref{betaAllorders} as conjectural.    In the Appendix we further discuss the status of the all-orders beta functions for 2D current-current perturbations in \cite{Gerganov}.



\section{Renormalization group trajectories}

\begin{figure}[t]
\centering\includegraphics[width=.5\textwidth]{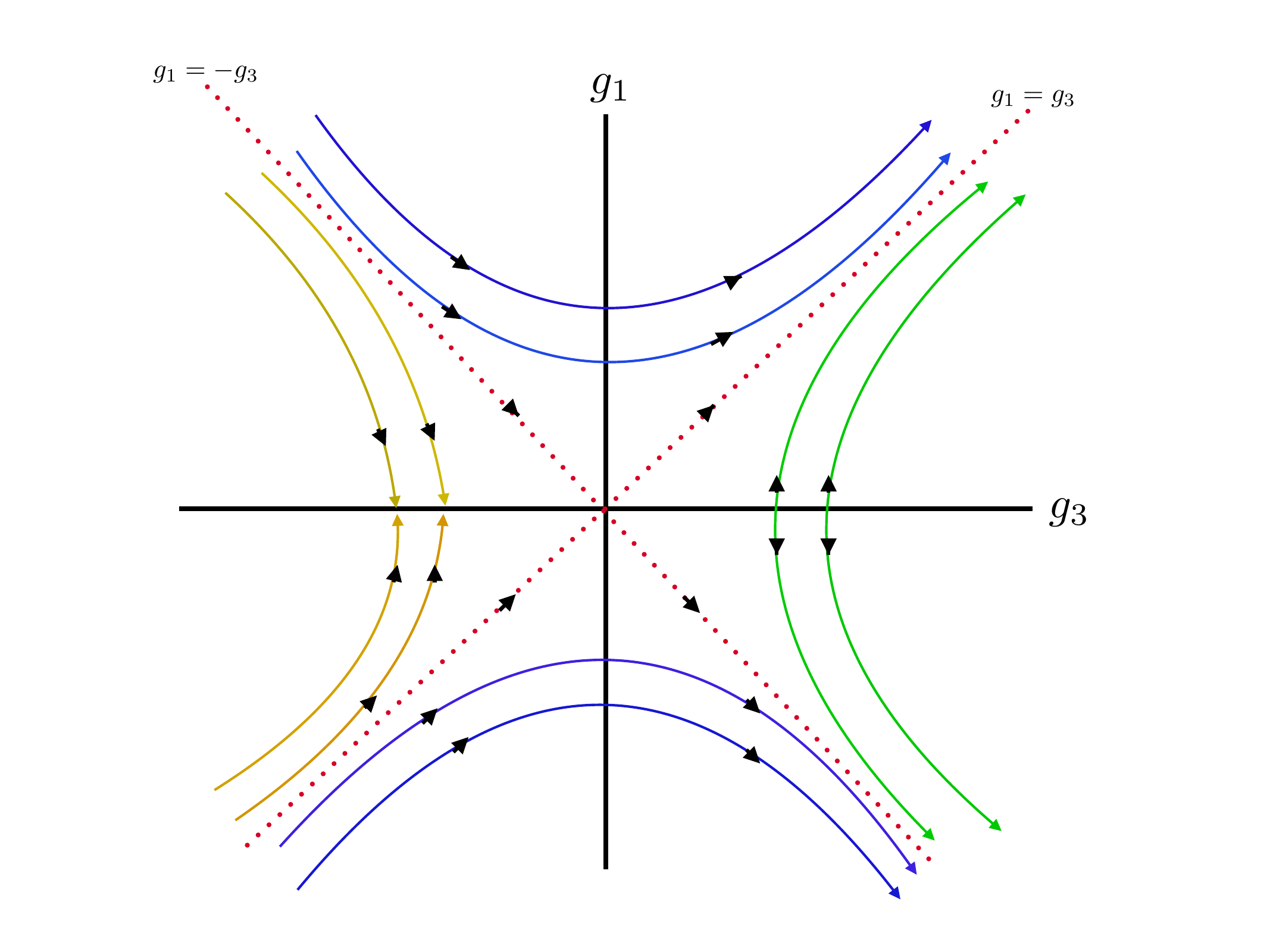}
\caption{RG flows to one loop. Arrows indicate the flow to low energies.}
 \label{Flows}
\end{figure}

In this section we map out the RG flows in the various regimes  of the couplings $g_1, g_3$ for the case of SU(2) broken to U(1).  We set $\kappa =1$ which is the correct value for our model.       In the 1-loop approximation,  the flows are shown in Figure \ref{Flows}.     At 1-loop the 
RG invariant $Q = g_1^2 - g_3^2$ implies the trajectories are hyperbolas.      Along the separatrices 
$g_3= \pm g_1$ the SU(2) is unbroken.    Here if $g_3>0$,  the theory is asymptotically free in the UV where $g_1 = g_3 = 0$.  
For $g_3<0$,  the theory is marginally irrelevant and flows to strong coupling in the UV wherein there is no known 
UV fixed point.   
Just below the separatrices where SU(2) is broken to U(1) there is a line of fixed points $g_1 =0$,   which can be marginally relevant or irrelevant.    Above the separatrices,     there are no fixed points anywhere,    and we will argue that this flow is 
actually cyclic \cite{ALDolls} (see below).      The cyclic flow is not at all rare for our model:    small deviations from the SU(2) invariant flows along the diagonal separatrices can go either either way:   either  they flow to the line of fixed points or do not and  are cyclic.       The aim of this section is to understand if this persists to higher orders based on the beta functions of the last section.     We will base our analysis on the geometric re-summation of the OPE diagrams of the last section,   which we argued leads to the all-orders beta functions \eqref{betaAllorders}.   These  beta functions are well defined in the limit $g \to \infty$,   which is especially important for the cyclic flows which require $g_3$ flowing to $\pm \infty$.   

The beta functions \eqref{betaAllorders} show that both $\beta_{g_1}$ and $\beta_{g_3}$ are zero when $g_1 =0$,   indicating a line of fixed points along the $g_3$ axis.     The anomalous scaling dimension  $\Gamma$ of the perturbation along this line is a function of $g_3$ and can be inferred from the slope of the beta function $\beta_{g_1}$.    
Generally speaking,  in 4 spacetime dimensions,   near a fixed point $g=g_c$ 
\beq
\label{betaGamma}
\beta_g = ( 4 - \Gamma ) (g - g_c)  .
\eeq
From the leading term in the series for  $ \beta_{g_1}$ about  $g_1=g_c=0$ one finds
\beq
\label{GammaFP}
\Gamma  (g_3) =  \frac{4}{1 + g_3 /4} .
\eeq
Based on this,   there are 3 distinct regions along the $g_3$ line of fixed points:

\bigskip
(i)    $g_3 >0$.    Here the perturbation is relevant with $0< \Gamma < 4$.   

\medskip
(ii)   $-4 < g_3 < 0$.     These are marginally irrelevant perturbations with $\Gamma > 4$. 

\medskip
(iii)  $g_3 < -4$.      These are still relevant perturbations but distinguished by  $\Gamma < 0$.

\def\gtilde{\tilde{g}}

\bigskip

It was shown in \cite{BLflow} that  RG flows  based on the beta functions \eqref{betaAllorders} can be extended to strong coupling $g_{1,3}  \to \infty$ where  the RG invariant $Q$ in \eqref{QAllorders} plays an essential role.     
This is due to a surprising strong-weak coupling duality
\beq
\label{duality1}
g_{1,3} \to \gtilde_{1,3} \equiv \frac{16}{ g_{1,3}}.
\eeq
Namely,   if $\gtilde = 16/g$,   then 
\beq
\label{duality}
\beta_{\gtilde}  (\gtilde ) \equiv \frac{\d \gtilde}{d g} \, \beta_g  = - \beta_g (g \to \gtilde) .
\eeq
The RG invariant $Q$ is also invariant under this strong-weak coupling duality:
\beq
\label{Qduality}
Q(\gtilde_1 , \gtilde_3 ) = Q(g_1, g_3) .
\eeq
This strong-weak coupling duality is completely un-anticipated,   and we don't know of any more fundamental reason to explain it besides by a posteriori  inspection of the
beta functions.    Even for the 2D current-current perturbations reviewed in the Appendix with identical beta functions,  the origin of this $g \to 16/g$ duality is not completely understood,    however  non-perturbative checks of it  based on the  Bethe ansatz confirm its validity.      For instance,   the action that defines the model does not have such a non-perturbative symmetry.    This situation is  analogous to the   
$b \to 1/b$ symmetry of the sinh-Gordon model with potential $\cosh (b \phi)$  where $b \to 1/b$ is a symmetry of the {\it S-matrix},   but not of the action.  (See the Appendix for how the sinh-Gordon model is realized as current-current perturbations in 2D.)    Based on analogy with the massless flows in 2D,    this strong-weak coupling duality 
is to some extent  analogous to $R \to 1/R$ duality of a free boson in 2D where $R$ is the radius of compactification.\footnote{See additional comments  in the Concluding Remarks.} 
It was also shown in \cite{BLflow} that it is consistent,  and required,   to endow the coupling constant space $(g_1, g_3)$ with the topology of a cylinder where one identifies $g_3 = \pm \infty$,   since $\beta(g_1, g_3) = \beta (g_1, - g_3) $ in the limit $|g_3| \to \infty$.   
This allows us to extend the flows to the entire $(g_1, g_3)$ plane,  including through the poles at $g_1 = \pm 4$,   which are self-dual points under \eqref{duality1}.       We refer the reader to \cite{BLflow} for more detailed explanations in the 2D case.  
  
\begin{figure}[t]
\centering\includegraphics[width=.9\textwidth]{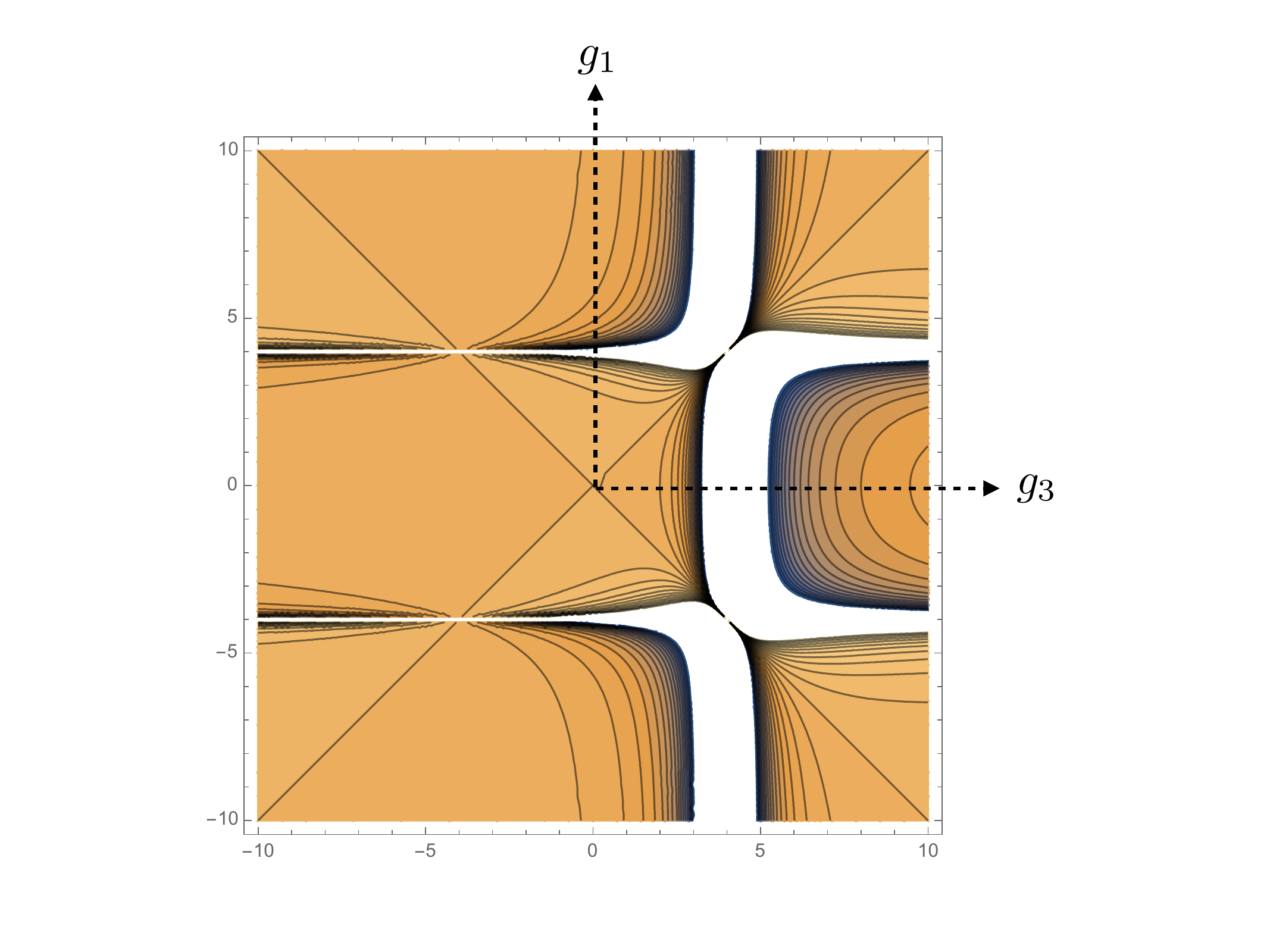}
\caption{Contour plot of the non-perturbative RG invariant $Q$  equation \eqref{QAllorders}. The points $g_3 = \pm \infty$ are identified.}
 \label{QAllContours}
\end{figure}

This leads a variety of RG flows which we now  itemize.     
 A contour plot of the non-perturbative $Q$ in \eqref{QAllorders}  provides a global picture of  the RG flows shown in Figure \ref{QAllContours}.
One sees that the set of all contours forms an interesting manifold.    Flows in the cyclic regime are shown separately in Figure 
\ref{QAllCyclic} since they are not clearly  distinguishable  in Figure \ref{QAllContours}.   One see's that they consistently cross the narrow bridges at the self-dual points.     

\subsection{Massive Flows with fixed points} 

\begin{figure}[t]
\centering\includegraphics[width=.6\textwidth]{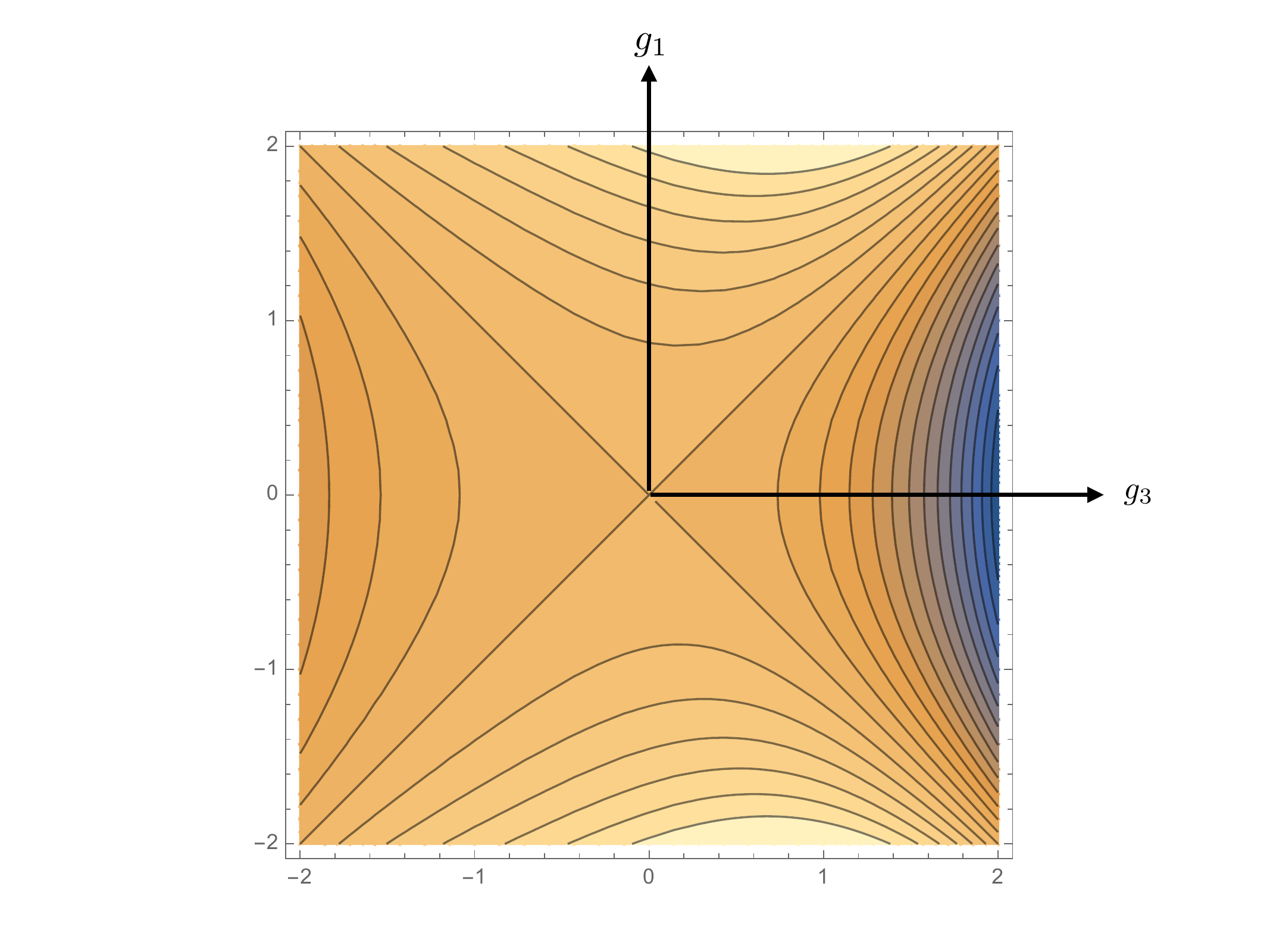}
\caption{Contour plot of the RG invariant $Q$ to 3-loops \eqref{Q3loop}.} 
 \label{3loopContours}
\end{figure}

For case (i) above,   namely $g_3>0$,   the perturbation is relevant  $0<\Gamma < 4$ and flows  to strong coupling in the IR.
We interpret this as a massive phase where the IR fixed point is empty.      In Figure \ref{3loopContours} we plot the constant $Q$ RG trajectories based only on the 3 loop result \eqref{Q3loop},  and one sees that the 1-loop flowchart  in Figure \ref{Flows}  is not significantly modified.    In comparison with the current-current perturbations reviewed in the Appendix,   this is a sine-Gordon like phase.

\def\UV{{\scriptscriptstyle {\rm UV}}}
\def\IR{{\scriptscriptstyle {\rm IR}}}

Next consider case (iii) above,  namely $g_3 < -4$.   These are also relevant perturbations,  with negative scaling dimension,  which are also expected to be massive.
$g_3 = -4$ is a self-dual point,   that is $\tilde{g_3} = g_3$.    One thus expects a different phase in this region.   In comparison with 2D,   this is analogous to a sinh-Gordon phase (see the Appendix).

\subsection{Massless Flows between two non-trivial fixed points for imaginary coupling} 

\begin{figure}[t]
\centering\includegraphics[width=.6\textwidth]{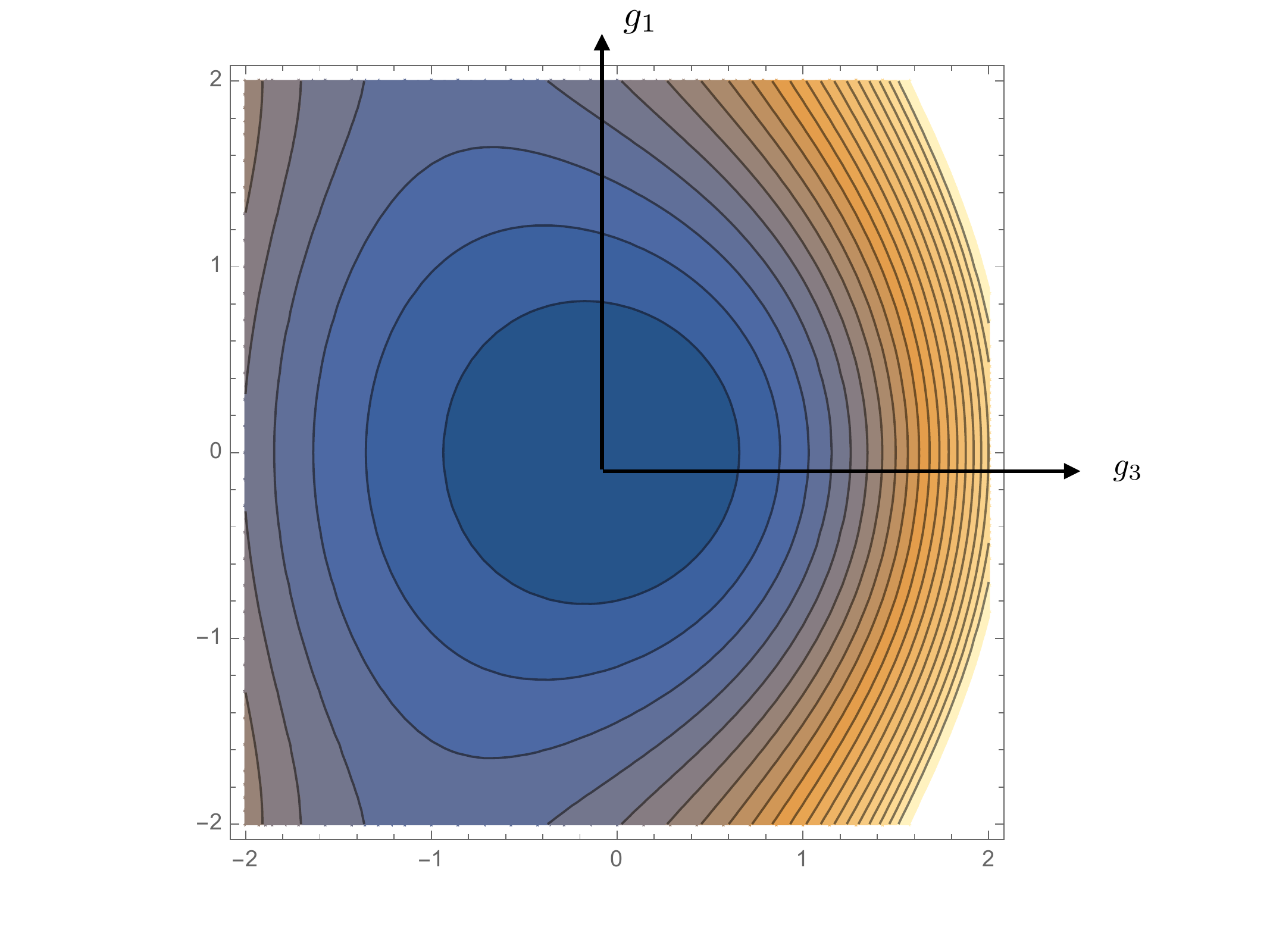}
\caption{Contour Plot of the RG invariant $Q$ for $g_1$ imaginary.   These flow contours begin and end along the critical 
$g_3$ line,  and do not pass through it as the above contours do.}
 \label{3loopContoursCircles}
\end{figure}

Massive theories typically terminate at an empty fixed point,  since at low energies the massive particles decouple.   
On the other hand,   flows that end  in the IR at a non-trivial CFT are expected to be massless,  since some massless degrees of freedom survive the flow.     There are no such flows in our models if $g_1$ and $g_3$ are real.   Since we are already considering non-unitary theories,   consider $g_1$ to be an imaginary coupling:
\beq
\label{imaginaryg}
g_1 \to i g_1 .
\eeq
Then the resulting beta functions still have real coefficients and thus make sense.       The RG invariant becomes 
\beq
\label{Qimag}
Q = -\frac{g_1^2 + g_3^2}{(1+ g_1^2/16)(1-g_3/4)^2}.
\eeq
Thus at small coupling,  the flows are no longer hyperbolas,  but circles,   and can thus can both originate and terminate along the line of fixed points $g_1 =0$.    
The trajectories based on the 3-loop $Q$  \eqref{Q3loop} are shown in Figure \ref{3loopContoursCircles}.   
 
This implies non-trivial flows between two fixed points along the $g_3$ axis.  
These flows start from a relevant perturbation with dimension $\Gamma_\UV<4 $ in the UV and arrive via an irrelevant perturbation in the IR with dimension $\Gamma_\IR>4$,  as they must.     An algebraic relation for these anomalous dimensions can be found using the RG invariant $Q$.   
When $g_1 =0$,   $Q=\[ g_3/(1-g_3/4) \]^2$.       
Equating 
\beq
\label{UVIR2}
Q_\IR = Q_\UV ~~~\Longrightarrow ~~~g_3^\IR =  \frac{2 g_3^\UV}{g_3^\UV-2}.
\eeq
Using the relation \eqref{GammaFP},   this can be expressed in terms of the $\Gamma$'s:\footnote{At 1-loop,   $g_\IR = - g_\UV$ and $\Gamma_\UV + \Gamma_\IR = 2\cdot D =8$.}
\beq
\label{GammaUVIR}
\Gamma_\IR = \frac{3 \Gamma_\UV -8 }{\Gamma_\UV -3} .
\eeq
 There is an interesting UV/IR duality,  in that the above equation also implies 
\beq
\label{GammaIRUV}
\Gamma_\UV = \frac{3 \Gamma_\IR  -8 }{\Gamma_\IR -3} .
\eeq
   A consistency check of the above formula is that when  $\Gamma_\UV = 4$,   then $\Gamma_\IR = 4$,  indicating no RG flow,   which corresponds to the point $g_1 = g_3 = 0$.    

This regime of massless flows requires  $\Gamma_\UV < 4$ and $\Gamma_\IR >4$.  
Based on \eqref{GammaUVIR},  this requires $\Gamma_\UV > 3$.     In terms of the coupling $g_3$ this corresponds to the regime  
\beq
\label{masslessregime}
3< \Gamma_\UV <4  ~~~~~ \Longrightarrow ~~~~ 0< g_3 < 4/3 .
\eeq
One sees that $g_3 = 4/3$ is special point where $\Gamma_\IR = \infty$.     We will say more about this in the next sub-section.

\def\Tbar{\bar{T}}

\subsection{Some special  rational  points in the  4D CFT's} 

We have already seen that the point $g_3 = 4/3$ is special since it is at the boundary of the massless flows where 
$\Gamma_\IR = \infty$.      One can surmise other potentially interesting special points as follows.     When 
$\Gamma_\IR = \Ddim$ where $\Ddim$ is an integer $\Ddim \geq 4$,   then the IR theory potentially  has some likely interpretations.
For instance,  if $\Ddim =8$,   this could represent massless flows that arrive to the IR via the operator 
$T_{\mu\nu} T^{\mu\nu}$,  $(T_\mu^\mu )^2$, or linear combinations,    where $T_{\mu\nu}$ is the dimension 4 energy-momentum tensor.   If this were the case,  
then this is analogous to $T\bar{T}$ perturbations of 2D CFT \cite{SmirnovZamo}.     For our model,   
$\Gamma_\UV = 16/5$ when $\Gamma_\IR = 8$,    which is rational and corresponds to $g_3 = 1$,  which is in the range \eqref{masslessregime}.     If one repeats this argument for the same beta function in 2 spacetime dimensions based on \eqref{GammaUVIR2D},  one finds 
 $\Gamma_\UV = 4/3$.    This turns out to be the scaling dimension of the perturbation for the  minimal $\CN =2$ supersymmetric model at Virasoro central charge $c=1$ \cite{BLsusy},  which is a special point of the sine-Gordon model at coupling $b^2 = 4/3$ in \eqref{SineGordonAction}.        Thus the 2D analog of this flow is such that it preserves the central charge $c$,   but the $\CN =2$ supersymmetry is broken in the flow and the remaining massless fields are goldstinos.\footnote{To our knowledge this observation is absent from the literature.}   Note that for such a flow in 2D,  the UV and IR CFT's are unitary,  however the RG flow is induced by non-unitary perturbations.   This flow is similar to massless flows that arrive in the IR to the CFT of a Majorana fermion via the dimension 4 operator $T \Tbar$ \cite{Alyosha},  where the UV central charge is $7/10$.         It was shown in \cite{UVIsing,AhnLeClair} that the only UV completions of flows that end at the Majorana description of the Ising model have UV central charge $c=7/10$ and $c=3/2$,  both of which have $\CN =1$ supersymmetry.\footnote{There are other flows that arrive  to the Ising model  via $T \Tbar$ based on  a  different spectrum 
 related to the $E_8$  Lie algebra \cite{AhnLeClair},  indicating multi-critical behavior.      In the notation in \cite{Tanaka},   this is a flow between $c<1$ minimal models  $M(11,12) \to M(3,4)$,  with central charge $c_\UV = 21/22$,  $c_\IR = 1/2$.      This flow is not included in \cite{Tanaka}  since there the focus on the generic 
 SU(2) quantum group modular fusion category of Verlinde lines,   whereas our $E_8$ based flow should be understood from the $E_8$ quantum group fusion category.}

 It is thus natural to consider massless flows that end with $\Gamma_\IR = \Ddim$ with $\Ddim$ an integer greater than $4$,  
 since the IR fixed point involves irrelevant operators with  integer dimensions and could correspond to Landau-Ginsburg theories where the scalar field $\phi$ has classical scaling dimension 1,  or have other interpretations such as 
 $T^2$ in the IR.        
The result is quite simple:
\beq
\label{special1}
\Gamma_\IR = \Ddim, ~~~~ \Longrightarrow ~~~\Gamma_\UV = \frac{3 \Ddim -8}{\Ddim-3}, ~~~~g_3^\UV = \frac{4(\Ddim -4)}{3\Ddim  -8} .
\eeq
This points to the existence of rational CFT's where important anomalous dimensions are rational numbers.   The latter is of course contingent on the  higher order beta functions \eqref{betaAllorders}.     For low $\Ddim \geq 4$ we collected some values in Table \ref{TableGammas}.

\begin{table}
\begin{center}
\begin{tabular}{||c|c||c|c||}
\hline\hline
$\Gamma_\UV$ &  $g_3^\UV$  & $\Gamma_\IR$    &   $g_3^\IR$    \\
\hline\hline 
$4$     &   $0$  &  $4$ &  $0$ \\
\hline
$\tfrac{7}{2} $     &   $\tfrac{4}{7} $  &  $5$ &  $-\tfrac{4}{5}$ \\
\hline
$\tfrac{10}{3} $     &   $\tfrac{4}{5} $  &  $6$ &  $-\tfrac{4}{3}$ \\
\hline
$\tfrac{13}{4} $     &   $\tfrac{12}{13} $  &  $7$ &  $-\tfrac{12}{7}$ \\
\hline
$\tfrac{16}{5} $     &   $1 $  &  $8$ &  $-2$ \\
\hline
$\tfrac{19}{6} $     &   $\tfrac{20}{19} $  &  $9$ &  $-\tfrac{20}{9}$ \\
\hline 
$\tfrac{22}{7} $     &   $\tfrac{12}{11} $  &  $10$ &  $-\tfrac{12}{5}$ \\
\hline 
$:$ & $:$ & $:$ & $:$ \\
$:$ & $:$ & $:$ & $:$ \\
\hline
$3 $     &   $\tfrac{4}{3} $  &  $\infty$ &  $-4$ \\
\hline\hline
\end{tabular}
\end{center}
\caption{Anomalous dimensions in UV and IR and the corresponding couplings $g_3$. The first row is the unperturbed free CFT.}
\label{TableGammas}
\end{table}

The special point $g_3 = 4/3$ appears as the limit when $\Ddim = \infty$.   From \eqref{special1},  
\beq
\label{special3}
\lim_{\Ddim\to \infty}  g_3^\UV = \frac{4}{3} .
\eeq
In this limit $\Gamma_\UV = 3$.      This implies that the coupling $g_1$ has dimension $1$ in the UV, i.e. it is a like a 
mass coupling.           One can argue that this special point is a free field theory,    since there is nothing to flow to if 
$\Gamma_\IR > \infty$,   thus there are effectively no flows.   Much more work would be needed to establish this.     However,   in support of this idea,  
let  us point out that for the models with  the same beta function in 2D reviewed in the Appendix   where \eqref{GammaUVIR} is replaced with \eqref{GammaUVIR2D},    there the analog of this special point is $\Gamma_\UV = 1$,   which is known to be a mass term for a free Dirac fermion in 2D.\footnote{Since relativistic fermion fields  $\psi$ have dimension $3/2$ in 4D,   this suggests that 
$g_3 = 4/3$ could  also be a  free fermion point,   where the perturbation is  $m \bar{\psi} \psi$ with $m$ identified by $g_1$.  
However  lacking a  formalism of bosonization in 4D,   this would be difficult to establish at this stage.}

\subsection{Cyclic RG flows}

Let us return to our original model with $g_1$ real.  
When $g_1^2 > g_3^2$,  one is above the SU(2) symmetric separatrices and the flows never encounter the $g_3$ line of fixed points at $g_1 =0$.      These are the most exotic flows,   since there are expectations  that all QFT's start and end at a fixed points,  at least for unitary theories.     
As discussed in the Introduction,   cyclic flows are consistent with various c-theorems if the model is non-unitary.    Furthermore,   in order to formulate such c-theorems
and their generalization to $4D$,   one needs to have a well-defined perturbation theory about {\it both}  the UV and IR fixed points \cite{Polchinski},    and such fixed points do not exist for the
cyclic flows we find.  
For our model,   if one is forced to understand all the flows in the $g_1, g_3$ coupling constant plane,   then cyclic flows in this regime are unavoidable.   

In this cyclic regime the RG invariant $Q >0$.   To 1-loop one can express the beta function $\beta_{g_3}$ in terms of $g_3$ and $Q$.   Integrating this,  the coupling constant as a function of the log of the length scale $\ell$ is 
\beq
\label{cyclic1}
g_3 (\ell) = \sqrt{Q} \tan \( \sqrt{Q}( \ell - \ell_0) \)
\eeq
where $\ell_0$ is an integration constant.  
The fundamental parameter of the theory is the  period $\lambda$ of the RG,   which is a simple function of $Q$ and thus an RG invariant: 
\beq
\label{cyclic2} 
g_3 (\ell + \lambda) = g_3 (\ell),     ~~~~~  \lambda = \frac{\pi}{\sqrt{Q}} .
\eeq
Thus one flows from $g_3 = -\infty$ to $g_3 = + \infty$ in a finite RG time $\lambda$.    Note that along the SU(2) invariant separatrices,   $Q=0$,   such that the RG period consistently goes to $\infty$.     

\begin{figure}[t]
\centering\includegraphics[width=.8\textwidth]{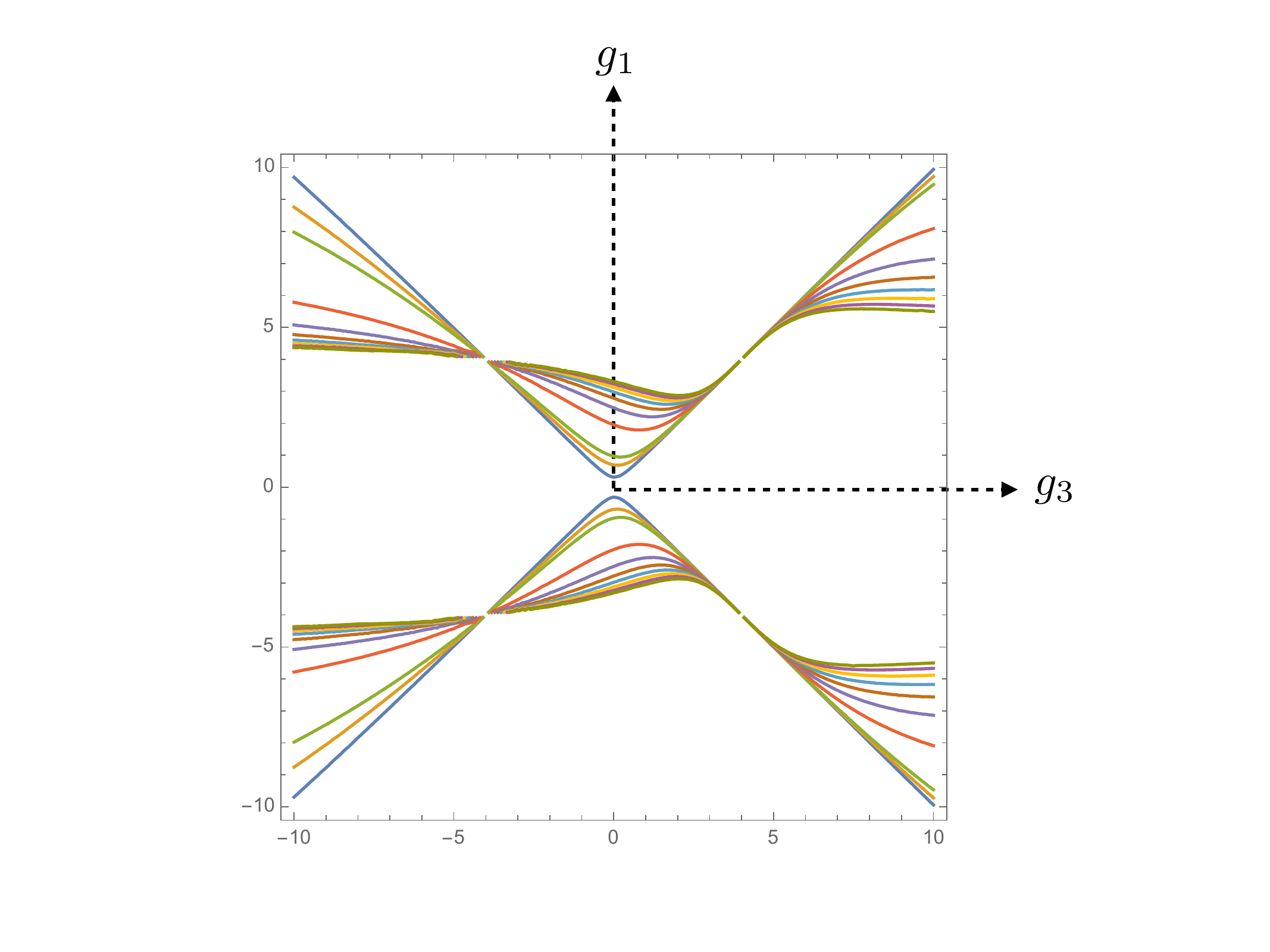}
\caption{Non-perturbative flows that are cyclic,  based on the contour plot of the non-perturbative RG invariant $Q$  in equation  \eqref{QAllorders} in the cyclic regime with
$0.1 < Q < 35$.  The points $g_3 = \pm \infty$ are identified such that the topology of coupling constant space is that of a cylinder. Nowhere does the RG trajectory reach the fixed points along the $g_3$ axis.}
 \label{QAllCyclic}
\end{figure}

 This cyclic  behavior could be spoiled at  higher orders in the couplings $g_1, g_3$  since the RG flows extend to $g_3 = \pm \infty$ which is beyond the 1-loop weak coupling regime where \eqref{cyclic2} was derived.   One can check that to 2-loops,  the period of the cyclic RG remains as in \eqref{cyclic2}.    The all-orders beta functions \eqref{betaAllorders} resolve this issue since these beta functions are well defined as $g\to \infty$.     These non-perturbative expressions simply lead to a doubling of  the RG period $\lambda$.    To see this,   we can eliminate $g_1$ from the beta function for $g_3$ using $Q$:
\beq
\label{g3Diffeq}
\beta_{g_3} = \frac{d g_3}{d \ell}  =  16 ~\frac{\( g_3^2 - 16 Q (g_3 -4)^2 \)\(1-Q(g_3 -4)^2 \)}{(g_3 + 4)^2 } .
\eeq
The above can be integrated and one still finds a cyclic RG \cite{LeClairSierra},  namely  $g_3 (\ell + \lambda) = g_3 (\ell)$,   where 
$\lambda$ is now  twice the 1-loop result:
\beq
\label{lambda2}
\lambda =  \int_{-\infty}^\infty  \frac{dg_3}{\beta_{g_3}} = \frac{2\pi}{\sqrt{Q}}.
\eeq
In Figure \ref{QAllCyclic} we plot the cyclic RG flows for a variety of positive $Q$ since they are difficult to discern  visually in 
Figure \ref{QAllContours}.    One sees that the flows smoothly  pass through the poles at the self-dual points $g_1,g_3 = \pm 4$.
This can be attributed to the fact that the flows approach the self-dual points with the correct slope,  namely along the 
SU(2) invariant flows along the diagonal.   This can be seen in  Figures  \ref{QAllContours},\ref{QAllCyclic} where flows cross the poles through very narrow bridges.      

\def\UV{{\scriptscriptstyle {\rm UV}}}
\def\IR{{\scriptscriptstyle {\rm IR}}}

\section{Concluding remarks}

To summarize,   we have computed the beta functions for our model  to 3 loops,   and showed that they do  not spoil  the main features of the RG flows based on  the 1-loop  approximation  found previously in \cite{ALDolls}.     For SU(2) broken to U(1) we re-summed an infinite number of contributions  and conjectured  a non-perturbative beta function.      This model has a line of fixed points,  which are new non-unitary CFT's in 4 spacetime dimensions.    There exists RG flows between non-trivial CFT  fixed points,  and we computed the relation between the anomalous dimensions of the perturbing operators in the UV and IR,   and furthermore identified some special points where these anomalous dimensions are rational.       There also exists a regime where the flows do not begin nor end at a fixed point but in fact are cyclic,  and we computed the period $\lambda$ of the  RG flow in terms of  the
RG invariant $Q$.     We argued that this circumvents the paradigm  that all QFT's begin or end at a fixed point as  proposed in  \cite{Schwimmer,Polchinski},  primarily  because the model is non-unitary.         

\bigskip
We close with discussion of some open questions raised by the above work,  of which there  are many.   

\bigskip
$\bullet ~$ 
The analysis of the higher order  RG flows in this paper points to the existence of some new non-unitary CFT's in 4 spacetime dimensions.    We computed some anomalous dimensions of relevant perturbing operators in the UV and  based on massless flows found some  special  rational exponents corresponding to flows that arrive in the IR via operators of integer dimension 
$\Gamma_\IR >4$,  such as $T^2$ where $T$ is the stress energy tensor in which case  $\Gamma_\IR = 8$ and $\Gamma_\UV = 16/5$.   
 Another interesting point is $\Gamma_\UV = 3$ which occurs at $g_3^\UV = 4/3$,  and we suggested that this could be a free fermion theory.   
  See Table \ref{TableGammas}.          If these calculations are correct,  they are still far from a complete understanding of these new CFT's and their flows.        For instance,      what are the primary fields and their correlation functions?   Is there a hidden symmetry that underlies our proposal
for the non-perturbative beta functions?       It is interesting to observe  that rational exponents can exist in both 2 and 4 spacetime dimensions,    whereas in 3 dimensions,   non-trivial exponents obtained by means of the epsilon expansion are highly non-perturbative and irrational \cite{WilsonFisher}.   This is perhaps due to there being no obvious analog to the 
$J \cdot  \Jtilde$ factorization  of marginal operators in our 4D model  to 3D,   which would require $J$ to have dimension $3/2$ such 
as $\phi^3$,  such that the marginal interaction goes as $\phi^6$.      In 2D the obvious generalization of this factorization is as a product of left/right moving chiral conserved currents, $J, \bar{J}$   which have dimension 1,   and here one can  formulate  perturbation theory of the WZW CFT using the OPE of these currents (see the Appendix).   
In this article,  we have generalized this OPE to 4D,  equation \eqref{JOPE},   and our proposed beta function is based almost entirely on this OPE.   However a comprehensive understanding the underlying  4D CFT,   analogous to the solution of the  WZW model in 2D,     has not been proposed in this article,   and is one of the main open avenues for further study.

\bigskip
$\bullet ~$ 
The operator $\K$ defined in equation \eqref{KU1}    was introduced  in order to obtain the  OPE \eqref{JOPE},   and the latter  required insertions of 
the $\K$-operator in correlation functions which modifies the inner-product on the Hilbert space according to equation \eqref{Knorm}.      Based on our results on massless flows above,   by analogy with the classification of 
massless flows between non-unitary CFT's in 2D \cite{Tanaka},  we suggest that $\K$ can be interpreted as a topological defect,   analogous to Verlinde lines in 2D CFT.  In the 4D context,   these should be codimension 2 defects,  namely two dimensional surfaces.   It is well accepted that such defects can modify the critical behavior in significant ways.     The massless flows we found along the critical line $g_1 =0$ should then be viewed as constrained by the U(1)  fusion category.   
 From this viewpoint,   the line of fixed points we found are analogous to the radius of compactification $R$  of a free boson in 2D,   which has a duality $R \to 1/R$  closely analogous to the strong-weak duality  $g \to 16/g$ of the beta functions,  equation \eqref{duality1}.
         The flows in \cite{Tanaka} are based on a study of the ${\rm SU(2)}_q$  quantum group modular fusion category where $q$ is a root of unity,   which is known to be a reduction of the U(1) category from various viewpoints,  such as quantum group reduction.

\bigskip
$\bullet ~$ 
The most unexpected result we found is that the beta function for our 4D model is identical to  that  of current-current interactions in 2D at least up to 3 loops in our specific renormalization scheme  based on the OPE.    Whereas the algebraic structure of the marginal perturbations is essentially the same as for 2D,    it is the fact that the higher order integrals,   which rely on the coefficients in \eqref{JOPE},  led to this result.    As stated above,  we had no a priori reason to expect this and have not provided 
distinct arguments for this to be the case.         The arguments of Section IVD led us 
 to propose the  non-perturbative beta functions \eqref{betaAllorders}.                
It would be very interesting if this feature can be extended to   a general construction  to ``lift"  2D CFT's to 4D.     A recent work along these lines is \cite{lifts}.

\def\SL2Z{{\rm SL}(2,\Z)}

\bigskip
$\bullet ~$ 
There is some evidence for an underlying $\SL2Z$ structure in our model.    Above we showed that the higher order beta functions have a $g \to 16/g$ duality,   like an S-duality $\tau \to -1/\tau$.     This strong-weak duality is an a posteriori  observation based on our computed beta functions and was unexpected,   thus we do not have an explanation of this from other distinct  non-perturbative arguments.        T-duality,  $\tau \to \tau + 1$ typically corresponds to a periodicity due to a $\vartheta$-angle,   as in  supersymmetric Yang-Mills theory.      
 The cyclicity of the RG required us to identify 
$g_3 = \pm \infty$,  endowing the 2-parameter coupling constant space $(g_1, g_3)$ with the topology of a cylinder,  which is a form of periodicity.   
 This by itself does not establish an ${\rm SL}(2, \Z)$ symmetry,   since we have not precisely identified the elliptic modulus 
 $\tau$ as a function of the 2 couplings.    
The role of $\SL2Z$ is perhaps  clearer for the fully anisotropic case with the 1-loop beta functions 
\eqref{1loopBetas} since,  as explained in Section IVA,    integration of the RG flow  already leads to elliptic functions \cite{Elliptic}. 
Furthermore the 2D models,  with identical beta functions to our models, were proposed to have the S-matrix of Zamolodchikov 
\cite{Zelip}  built out of elliptic functions \cite{Elliptic}.    This S-matrix is characterized by a $\Z_4$ symmetry which is a subgroup of $\SL2Z$,  
 generated by the matrix  $A =  -i \sigma^2 = \begin{pmatrix} 0 & -1 \\ 1 & 0 \end{pmatrix} $.
Furthermore,   it is known that this S-matrix transforms covariantly under $\SL2Z$, 
meaning the S-matrix transforms with phase factors or gauge transformations that preserve the structure of the Yang-Baxter equation \cite{Baxter,Belavin}.  
In \cite{Elliptic},  for  the SU(2) broken to U(1) model,  which is  the cyclic sine-Gordon model reviewed in the Appendix,  the latter  can be obtained from certain limits of Zamolodchikov's elliptic S-matrix,  and this is perhaps a way of understanding what is left of the $\SL2Z$ structure for our 
two parameter model.

\bigskip
$\bullet ~$ 
For the cyclic flows we have found,   it's important to understand the physical manifestations of this cyclicity  as far as the  spectrum of particles and their scattering. 
For some 2D models with identical beta functions,  such as the cyclic sine-Gordon model \cite{LeClairSierra},   the cyclicity is manifested as an infinite number of resonance poles in the S-matrix with Russian Doll scaling as in equation \eqref{ERussianDoll}.   It would be very interesting to explore similar manifestations in 4D.


\section{Acknowledgements} 

We wish to  thank  Luis Alvarez-Gaum\'e, Denis Bernard,  Igor Klebanov,  Cheng-Yang Lee,     Michael Peskin,  Germ\'an Sierra and 
Takahiro Tanaka  for discussions.     We also wish to thank the late Ken Wilson for correspondence on  his last works with Glazek \cite{GW1,GW2},  
around the time  we published \cite{BLflow}.

\section{Appendix:  Comparison with 2D current-current perturbations}

It is instructive to compare the above  results for our 4D model  with results for some  exactly solvable models in 2D    which have a very similar Operator Algebra (OPE's)  for the marginal perturbations we introduced. 
For instance,   this  can provide  insights into  what are  the physical manifestations of cyclic flows in this context of 4D QFT.

In \cite{Gerganov} an all-orders beta function was proposed for the most general case of anisotropic current-current perturbations of Wess-Zumino-Witten models of 2D CFT for an arbitrary Lie group \cite{WittenWZW,KZ}.\footnote{The paper 
\cite{Gerganov} also included  super Lie groups,  the latter being motivated by applications to disordered systems and Anderson localization.}       
The models are defined by the action 
\beq
\label{WZW}
S = S_{\rm WZW}  +  \sum_{A,a,b}    g_A  \, d^A_{ab} \, \int d^2 x \,  J^a (x) \bar{J}^b  (x) , 
\eeq
where $J^a$,  $\bar{J}^a$ are the left,  right  chiral conserved currents of the WZW model,  $d^A_{ab}$ are  fixed bi-linear coefficients,  and $g_A$ are couplings.    As in the present article,    the prescription for the beta-functions was also based on the OPE of the currents $J^a$.    In the conformal WZW model,
the currents $J^a, \bar{J}^a $ are functions of $z = x_1+i x_2$ and $\zbar = x_1-i x_2$ respectively,  where $x_1,x_2$ are the two euclidean spacetime coordinates.   They satisfy the OPE   
\beq
\label{CurrentOPE}
J^a (z) J^b (0) = \frac{k \delta^{ab}}{2 z^2} + \inv{z}  f^{abc} J^c (0) + \ldots
\eeq
and similarly for $\bar{J} (\zbar )$.   This should be compared with the OPE \eqref{JOPE}.      Above,  $k$ is a fundamental parameter,   the level of the affine Lie algebra,   which is a positive integer for unitary 
theories.\footnote{The conventions for the normalization of the 2D currents $J^a$ is different in minor ways, by factors of $\sqrt{2}$,   from the definitions of the operators $J^a$ in the body of this paper.   In this appendix,  we do not pay close attention to overall factors,   since for purposes of comparison with our 4D models,   we normalize the couplings such that the 1-loop beta functions agree.}
For $\SU2$,  the coefficients $d^A_{ab}$ can be chosen such that 
\beq
\label{perturbation} 
\sum_{A,a,b=1,2,3}   g_A  \, d^A_{ab}  \,\,J^a \bar{J}^a = g_1 \(J^+ {\bar J}^- + J^- {\bar J}^+ \) + g_3 J^3 \bar{J}^3 ,
\eeq
where  $J^\pm = J^1 \pm i J^2 $.      
The beta-functions computed in \cite{Gerganov} are precisely those in equation \eqref{betaAllorders} with $\kappa =k$.    For the SU(2) symmetric case $g_1 = g_2$,   the beta function agrees with the one proposed  by Kutasov \cite{Kutasov}.   

When the level $k=1$,  the currents can be bosonized in terms of the left/right components of a single scalar field 
$\phi = \varphi (z) + \bar{\varphi} (\zbar)$:
\beq
\label{Bosonization}
J^\pm   =   \exp \( \pm i \sqrt{2} \,\varphi \)  , ~~~~~J^3 = i  \d_z \varphi .
\eeq
The action becomes 
\beq
\label{CurrentAction}
S = \inv{4\pi} \int d^2 x \( \half (\d \phi)^2 + g_1 \cos (\sqrt{2} \phi ) + g_3 (\d \phi )^2 \) .
\eeq
The $g_3$ term can be incorporated into the kinetic term,  and by rescaling the scalar field,  one obtains the sine-Gordon model:
\beq
\label{SineGordonAction}
S = \inv{4\pi} \int d^2 x \( \half (\d \phi)^2 + g_1 \cos (b \phi ) \) ,
\eeq
where $b$ is a function of the couplings $g_1, g_3$  that can be found in \cite{BLflow}.   Henceforth we specialize to $k=1$.

Using the strong-weak coupling dualities \eqref{duality} and \eqref{Qduality} it was shown in \cite{BLflow} that the flows can be completed to strong coupling.    There are essentially 3 regions of couplings with different RG behavior depending on the value of $Q$.         The main features are the following:

\medskip
$\bullet$  When $Q$ is negative with $|g_3| < 4$,  the coupling $b$ in \eqref{SineGordonAction} is real.   This is a phase corresponding to marginally relevant or irrelevant perturbations of the sine-Gordon model.     

\medskip
$\bullet$   When $Q$ is negative with $|g_3| > 4$, $b$ is imaginary.    This  phase corresponds to the  sinh-Gordon model where 
$\cos(b \phi)$ becomes $\cosh (b \phi)$.   

\medskip
$\bullet$    When $Q$ is positive,      the RG is cyclic,   $g_3 (\ell + \lambda) = g_3 (\lambda)$,     with period $\lambda =2 \pi/\sqrt{Q}$.     The S-matrix was proposed to be the analytic continuation of the usual sine-Gordon model \cite{ZamoZamo}  to the appropriate value of $b$ \cite{LeClairSierra}.    The resulting S-matrix  has an infinite number of resonances with masses $m_n$ with Russian Doll scaling behavior 
\beq
\label{resonantmass}
m_n = 2 M_s \cosh (n \lambda/2 ) \approx M_s  e^{n \lambda/2}  ~~~{\rm as ~ } n \to \infty 
\eeq
where $n$ is an integer and $M_s$ the soliton mass 
\cite{LeClairSierra}.   

\medskip
The anomalous dimensions $\Gamma (g_3)$ along the critical line can be found from 
\beq
\label{GammaOfg3A}
\beta_{g_1}  = (2 - \Gamma(g_3)  ) g_1 + \ldots .
\eeq
This gives 
\beq
\label{GammaOfg3}
\Gamma (g_3) =   \frac{2(4-g_3)}{(4 + g_3 )} .
\eeq

$\bullet$    When $g_1$ is imaginary,  i.e. $g_1 \to i g_1$,    then this is the so-called imaginary sine-Gordon model \cite{Fendley1,Fendley2},   which has massless flows between two fixed points on the critical line $g_1 =0$.    These flows do not change the Virasoro central charge $c=1$,   and the UV and IR fixed points are those of a free boson with different compactification radius.   That the central charge does not change already violates the c-theorem \cite{ctheorem},   presumably because of the non-unitarity.    Quantum Group reductions of these flows describe flows between unitary minimal models with 
$c<1$ \cite{Fendley2}.         Repeating the arguments of Section VB  based on the beta functions \eqref{betaAllorders},   the relation between the anomalous dimensions in the UV and IR is now in 2D  
\beq
\label{GammaUVIR2D}
\Gamma_\IR = \frac{\Gamma_\UV}{\Gamma_\UV -1} .
\eeq
These flows only exist for $1<\Gamma_\UV <2$,    where $\Gamma_\UV =1$ corresponds to the free-fermion point of the sine-Gordon model.   This model is integrable and the relation \eqref{GammaUVIR2D},  which was derived only from the beta functions \cite{BLflow},   agrees exactly  with the thermodynamic Bethe-ansatz analysis in \cite{Fendley2},  and this  provides a non-trivial check of the validity of our beta functions to all orders in 2D.      

Let us make a few comments on the status of the conjectured all-orders beta functions in equation \eqref{betaAllorders} in the 2D context of this Appendix.       The fact that these beta functions lead to the result \eqref{GammaUVIR2D} which agrees with the fully non-perturbative Bethe Ansatz results in \cite{Fendley1,Fendley2}  is the strongest indication of their validity.   
Another strong positive check  is that  the  all orders beta function in \cite{Gerganov} correctly predicted the RG period $\lambda$ based on comparison with the 
exact S-matrix of the cyclic sine-Gordon model,  where as previously discussed the RG period $\lambda$ determines the poles for the Russian Doll resonances \cite{LeClairSierra}.     On the other hand,  
  in \cite{LudwigWiese} it was argued that for 
Thirring-type models there is potentially a problem with the all-orders conjecture \eqref{betaAllorders}  that arises at 4 loops.  
This was based on studying the peculiar  $k=0$ case,  which is not physically realized in our specific  models.   
   These calculations were performed  with Feynman diagrams rather than just the OPE of the currents.    If these 4-loop corrections were correct,   they would spoil the agreement with the exact Bethe ansatz for the physical $k\neq 0$ theories.     Let us make the following additional comments.    As mentioned above,  beyond 2 loops the beta functions are not universal but are scheme dependent,  and $k=0$ is not a generic physical case realized in our models.       We can present two concrete counter arguments to the criticism in \cite{LudwigWiese}. 
First,     for applications to disordered systems,   in models where the $gl(N|N)$ 
super-algebra is relevant \cite{Guru},   an all-orders beta function was computed based on very different arguments than  summing the perturbative OPE arguments involving the currents $J^a$ as was done in \cite{Gerganov}.      This was possible since the theory considered is almost conformal,   such that the beta function could be determined by various field redefinitions \cite{Guru}.     The beta function computed in \cite{Guru} for the $gl(N|N)$ super-current algebra agree exactly with the general formulas in \cite{Gerganov},  where the latter was based on the OPE considerations in this paper.  
Second,   subsequently new kinds of integrable sigma models were discovered where a comparison with the all-orders beta functions in 
\cite{Gerganov} could be made,   and positive agreement was found \cite{Sfetsos}.      The latter suggests that in a particular renormalization group scheme the beta functions in \cite{Gerganov} are valid.      In summary,   although these issues have not been fully resolved rigorously,  there are enough checks on the beta functions \eqref{betaAllorders} to justify their validity at this stage,  at the very least to understand the global structure of RG flows.


\end{document}